\documentclass[
notitlepage,%
aip,
jmp,
a4paper,%
final,%
reprint,%
]{revtex4-1}

\usepackage{amsmath,amsfonts,amsthm,amssymb,times,mathrsfs,txfonts}
\usepackage{float}
\usepackage{bm}% bold math

\let\Re\relax
\DeclareMathOperator{\Re}{Re}
\let\Im\relax
\DeclareMathOperator{\Im}{Im}
\newcommand{\field}[1]{\mathbb{#1}}
\newcommand{\fs}[1]{\mathsf{#1}}
\newcommand{\ABC}{\text{ABC}}

\DeclareMathOperator{\sgn}{sgn}
\newcommand{\laplace}{\mathscr{L}}
\newcommand{\fourier}{\mathscr{F}}
\renewcommand{\binom}[2]{\begin{pmatrix}#1\\#2\end{pmatrix}}
\newcommand{\tp}{\intercal}% transpose operation
\DeclareMathOperator*{\supp}{supp}

\newcommand{\vv}[1]{\mathbf{#1}}
\newcommand{\vbs}[1]{\boldsymbol{#1}}

\newcommand{\Schwartz}{\mathsf{S}}
\newcommand{\OP}{\textrm{OP}}
\newcommand{\et}{\textit{et~al.}}

\usepackage[pdftex]{graphicx}
\graphicspath{{./figs/}}
\usepackage[caption=false]{subfig}
\usepackage[pdftex,colorlinks=true,bookmarks=true,citecolor=blue,urlcolor=blue]{hyperref} %pdflatex
\usepackage{enumitem}
\newtheorem{rmk}{Remark}[section]

\begin{document}

%\preprint{}

\title{On the nonreflecting boundary operators for the general two 
dimensional Schr\"odinger equation}


\author{Vishal Vaibhav}
\email[]{vishal.vaibhav@gmail.com}
%\thanks{}
\noaffiliation{}
%\affiliation{Delft Center for Systems and Control, Delft University of Technology, 
%Mekelweg 2, 2628 CD Delft, The Netherlands}% <-this % stops a space

\begin{abstract}
Of the two main objectives we pursue in this paper, the first one consists in the studying
operators of the form $(\partial_t-i\triangle_{\Gamma})^{\alpha},\,\,\alpha=1/2,-1/2,-1,\ldots,$
where $\triangle_{\Gamma}$ is the Laplace-Beltrami operator. These operators arise in the context
of nonreflecting boundary conditions in the pseudo-differential approach for the
general Schr\"odinger equation. The definition of such operators is discussed in 
various settings and a formulation in terms of fractional operators is provided. The 
second objective consists in deriving corner conditions for a rectangular domain
in order to make such domains amenable to the pseudo-differential approach. Stability
and uniqueness of the solution is investigated for each of these novel boundary
conditions.
\end{abstract}

%\keywords{}

\maketitle

\section{Introduction}
In this article, we consider the problem of construction of nonreflecting boundary
condition for the general two dimensional Schr\"odinger equation. In particular,
we consider the following initial value problem (IVP):
\begin{equation}
\begin{split}\label{eq:2D-NLS}
&i\partial_tu+\triangle u+\phi(\vv{x}, t, |u|^2)u=0,\quad(\vv{x},t)
\in\field{R}^2\times\field{R}_+,\\
&u(\vv{x},0)=u_0(\vv{x}),\quad\vv{x}\in\field{R}^2.
\end{split}
\end{equation}
where $\field{R}_+$ denotes the non-negative real numbers ($\field{R}$) and the 
initial data is assumed to be supported within the computational domain,
$\Omega_i$, i.e., $\supp\,u_0\subset\Omega_i$. Let the boundary of the computational domain 
be denoted by $\Gamma$. Further, we assume that the potential function, $\phi$,
is real-valued. This problem has been treated by several 
authors~\cite{Menza1996,Menza1997,AB2001,S2002,HH2004,ABM2004,Zheng2007,FP2011,ABK2012,ABK2013}.
Exact formulations of the \emph{transparent boundary condition} (TBC) for the free 
Schr\"odinger equation on convex domains, $\Omega_i$, with smooth boundary was provided by
Sch\"adle~\cite{S2002} in terms of a single and a double layer potential. The
special case of a circular domain was treated by Han and Huang~\cite{HH2004}.
Earlier attempts to derive an exact TBC for a rectangular domain by Menza proved to 
be problematic on account of the presence of corners~\cite{Menza1996,Menza1997}. This 
problem is resolved in a recent work by Feshchenko and Popov~\cite{FP2011}. For
the spatially discretized free Schr\"odinger equation, TBCs on a rectangular domain have 
also been recently reported by Ji~\et~\cite{JPA2018} where exact form of the
Green's function was obtained on a purely discrete level to construct the
\emph{discrete} TBCs. The stability analysis of these discrete TBCs is carried out in 
Ref.~\onlinecite{JYAT2018}. On account of the lack of integrability, these techniques cannot be applied to 
the general Schr\"odinger equation and one has to turn to approximate methods
(see Refs.~\onlinecite{CiCP2008,ALT2017} for a comprehensive literature survey).

For the approximate methods, we restrict ourselves to the pseudo-differential approach 
(in particular, the gauge transformation strategy~\cite{ALT2017}) for constructing approximate 
nonreflecting boundary conditions referred to as 
the \emph{absorbing boundary conditions} or \emph{artificial boundary conditions} 
for various types of computational domains. Our goal is to
understand operators of the form
$(\partial_t-i\triangle_{\Gamma})^{\alpha},\,\,\alpha=1/2,-1/2,-1,\ldots,$ 
which appeared in the works of Menza~\cite{Menza1996,Menza1997} and 
Antoine~\et~\cite{ABM2004,ABK2012}. Several aspects of such artificial boundary
conditions (ABCs) which comprises these operators are not quite well 
understood; we discuss these issues which motivate
the present work in the subsequent paragraphs. 

Contrary to the existing belief that the aforementioned operators can only be 
implemented via a Pad\'e approximation~\cite{BGM1981} of a monomial of fractional 
degree $\alpha$, i.e., $z^{\alpha}$, it is shown in various settings that this 
operator can be expressed in terms of fractional
operators. For arbitrary functions $f(x,t),\,\,(x,t)\in\field{R}\times\field{R}_+$, the operation 
$(\partial_t-i\triangle_{x})^{\alpha}f(x,t),\,\,\alpha=1/2,-1/2,-1,\ldots,$ requires the knowledge of the
function over its entire support. If $\supp_xf(x,t)\subset\Gamma_x,\,\,\forall t\geq0$ or $f(x,t)$ is
periodic with respect to $x$ and it can be uniquely defined by its values at
$x\in\Gamma_x$ for all $t\geq0$, then it is shown that a formulation particularly convenient for
expressing the TBCs/ABCs for the IVP in~\eqref{eq:2D-NLS} can be 
developed\footnote{The TBCs derived by Feshchenko and Popov~\cite{FP2011}
happen to be a special case where this operation can be carried without
having to consider the function $f(x,t)$ over its entire support. The function 
$f(x,t)$ here refers to the restriction of the field
$u(\vv{x},t),\,\vv{x}\in\field{R}^2,$ to the segments of the boundary of the 
rectangular domain.}. Let us remark that the numerical implementation of 
such operators is not being presented in this paper; however, it can be easily seen that
the formulation developed in this paper makes these operators amenable to
\emph{convolution quadrature}~\cite{Lubich1986,*Lubich1988I,*Lubich1988II,*Lubich1994}. It might 
be expected that the new scheme affords improvement in 
accuracy over the existing Pad\'e approximation based method\footnote{A distinction must be made between 
the Pad\'e approximation based methods which
are applied to the operator $(\partial_t-i\triangle_{x})^{\alpha}$ and
those applied to the fractional operator formulation of this operator discussed
in this paper.} reported 
in Refs.~\onlinecite{MZB1977,ABK2013}.

Further, it is well known that the pseudo-differential approach cannot be applied to
computational domains with corners. This precludes the rectangular domain which happens to be
a very convenient choice of the computational domain. The ABCs involving the
operators of the form $(\partial_t-i\triangle_{\Gamma})^{\alpha}$ cannot be adapted
to the rectangular domain in a straightforward manner; however, the ABCs obtained as local approximations
(with respect to $x$) or, equivalently, the high-frequency approximations 
of this operator admit of the possibility of constructing the so-called 
\emph{corner conditions}. We demonstrate this possibility for the free as well as the
general Schr\"odinger equation given by~\eqref{eq:2D-NLS}. Our approach is closely related 
to the ideas presented in Refs.~\onlinecite{BJR1990,Vacus2005}. 

Another program that we have followed in this paper is of obtaining the
well-known ``energy'' estimate which can be introduced as follows: In the
context of electromagnetic fields, the square of the
$\fs{L}^2$-norm is related to the total energy, $E(t)$, of the field given by
\begin{equation}
E(t;\Omega)=\int_{\Omega}|u(\vv{x},t)|^2d^2\vv{x}=\|u(\cdot,t)\|^2_{\fs{L}^2(\Omega)},
\quad\forall t\geq0.
\end{equation}
The total energy, $E(t;\field{R}^2)$, remains constant in a conservative system
which is also true of~\eqref{eq:2D-NLS} given that $\phi$ is real-valued. If the initial field is 
supported in $\Omega_i$, the fact that the energy content of the 
field in $\Omega_i$, at any later time $t>0$, cannot exceed that of the initial field is expressed 
by the inequality $E(t;\Omega_i)\leq E(0;\Omega_i)$ or, equivalently,
\begin{equation}
\|u(\cdot,t)\|_{\fs{L}^2(\Omega_i)}\leq\|u_0\|_{\fs{L}^2(\Omega_i)},
\quad\forall t>0,
\end{equation}
for the IVP in~\eqref{eq:2D-NLS} with TBCs/ABCs as boundary conditions involving
operators of the form $(\partial_t-i\triangle_{\Gamma})^{\alpha}$ or high-frequency
approximations of it\footnote{The corner conditions are also incorporated in
this equivalent formulation.}. Under the assumption that the solution exists, this 
inequality guarantees the stability as well as the uniqueness of the solution of the
equivalent initial boundary-value problem (IBVP). In 
certain cases, this result can be obtained by resorting to a general 
construct such that for any pseudo-differential operator, $P$, and a function
$u\in\fs{C}_0^{\infty}(\Omega)$ we have
\begin{equation}
2\Re\langle u|P u\rangle=\langle u|P u\rangle+\langle u|P^{\dagger} u\rangle,
\end{equation}
where $P^{\dagger}$ is adjoint of the operator $P$, $\Re$ stands for real part 
and $\langle u|v\rangle=\int_{\Omega}u^*vd\Omega$. Given the symbol
$\sigma_P$ of $P$, the symbol of the adjoint, $\sigma_{P^{\dagger}}$, can be computed 
using the following general formula: Let $\field {N}^n$ denote the 
$n$-ary Cartesian power of set of non-negative integers. Assuming 
$y\in\field{R}^n$ and $\zeta$ the covariable of $y$, we have
\begin{equation}\label{eq:symbol-adj}
\sigma_{P^{\dagger}}(y,\zeta)\sim\sum_{\alpha\in\field{N}^n}
\frac{1}{\alpha!i^{|\alpha|}}
\partial^{\alpha}_{\zeta}[\partial^{\alpha}_{y}\sigma^*_P(y,\zeta)],
\end{equation}
where $\sigma_P^*$ stands for complex conjugate of $\sigma_P$ 
(complex conjugate of $z\in\field{C}$ is also denoted by $\overline{z}$). A more detailed 
discussion of this approach is provided in the Appendix~\ref{app:psido}. Let us
note that~\eqref{eq:symbol-adj} determines $\sigma_{P^{\dagger}}$ as an
asymptotic series; therefore, 
the energy estimates obtained by retaining only
the leading order terms in~\eqref{eq:symbol-adj} holds 
only in a asymptotic sense. Given that it is
usually a formidable task to demonstrate the stability of the IBVP involving
TBCs/ABCs, it appears to be a more realistic goal to establish this ``weak''
form of stability\footnote{We hope that this approach can also be extended to ABCs
obtained for the 1D case using microparametrices~\cite{Lascar1977} in 
Refs~\onlinecite{V2011,V2014}. Further, there are more general type of operators proposed
in Ref.~\onlinecite{ABK2012} that do not appear to be amenable to exact analysis. 
These possibilities will be explored in a future publication.}.

The discussion of the primary results in this paper is broadly divided into two
sections: Section~\ref{sec:fse} deals with the free Schr\"odinger
equation while Section~\ref{sec:gse} deals with the general Schr\"odinger
equation. For each of these problems, we consider two types of domains, namely,
the rectangular domain (or, infinite strip with periodic boundary condition along the
unbounded direction) and convex domains with smooth boundary. The basic definition of 
the operator $(\partial_t-i\triangle_{\Gamma})^{\alpha},\,\,\alpha=1/2,-1/2,-1,\ldots,$ 
is discussed in various settings. Two families of ABCs are considered in this
paper: first one obtained via the standard pseudo-differential approach and the second one
obtained as the high-frequency approximation of the former. The derivation of
corner conditions and the study of stability and uniqueness of the solution are
carried out separately for each of these problems in the subsections.
%%%%%%%%%%%%%%%%%%%%%%%%%%%%%%%%%%%%%%%%%%%%%%%
\section{Free Schr\"odinger equation}\label{sec:fse}
Let us start our discussion with the linear case of the 
equation~\eqref{eq:2D-NLS} with null-potential, i.e.,
\begin{equation}\label{eq:2D-SE}
\begin{split}
&i\partial_tu+\triangle u=0,\quad(\vv{x},t)\in\field{R}^2\times\field{R}_+,\\
&u(\vv{x},0)=u_0(\vv{x}),\quad\vv{x}\in\field{R}^2.
\end{split}
\end{equation}
The problem of constructing transparent boundary condition can be exactly treated for the case of 
compactly supported initial data for a computational domain $\Omega_i$ with a general 
smooth boundary~\cite{S2002}, say, $\Gamma$. The basic approach involves solving the 
initial boundary value problem on the exterior domain which is defined by 
$\Omega_{e}=\field{R}^2\setminus\overline{\Omega}_i$. It is known that a boundedness 
condition at infinity does not ensure that a unique solution of the
IVP in~\eqref{eq:2D-SE} exists; one additionally needs to impose a Sommerfeld-like 
radiation condition at infinity in order to exclude all the incoming waves from 
infinity. This condition reads as
\begin{equation}
\lim_{|\vv{x}|\rightarrow\infty}
\sqrt{|\vv{x}|}\left(\nabla u\cdot\frac{\vv{x}}{|\vv{x}|}
+e^{-i\frac{\pi}{4}}\partial_t^{1/2}u\right)=0.
\end{equation}
Using the decomposition of $u(\vv{x},t)\in
\fs{L}^2(\field{R}^2)=\fs{L}^2(\Omega_i)\oplus \fs{L}^2(\Omega_e)$ 
and introducing the fields $v(\vv{x},t)$ and $w(\vv{x},t)$ we have
\begin{equation}
\begin{split}
&\left\{\begin{aligned}
\label{decomposition}
&i\partial_tv+\triangle v=0,\quad(\vv{x},t)\in\Omega_i\times\field{R}_+,\\
&v(\vv{x},0)=u_0(\vv{x}),\quad\vv{x}\in\Omega_i;\\
&\\
&i\partial_tw+\triangle w=0,\quad(\vv{x},t)\in\Omega_e\times\field{R}_+,\\
&u(\vv{x},0)=0,\quad\vv{x}\in\Omega_e,\\
&\lim_{|\vv{x}|\rightarrow\infty}\sqrt{|\vv{x}|}
\left(\triangle
w\cdot\frac{\vv{x}}{|\vv{x}|}+e^{-i\frac{\pi}{4}}\partial_t^{1/2}w\right)=0;
\end{aligned}\right.\\
&v(\vv{x},t)|_{\Gamma}=w(\vv{x},t)|_{\Gamma},\,
\partial_{n}v(\vv{x},t)|_{\Gamma}=\partial_{n}w(\vv{x},t)|_{\Gamma}.
\end{split}
\end{equation}
Construction of the nonreflecting boundary conditions involves solving the exterior
problem exactly. In the following sections, we first consider a infinite strip 
as computational domain (or periodic boundary condition in the direction it
extends to infinity) then extend the results to a rectangular domain. 

\begin{figure}[!htbp]
\begin{center}
\def\myscale{1}
\includegraphics[scale=\myscale]{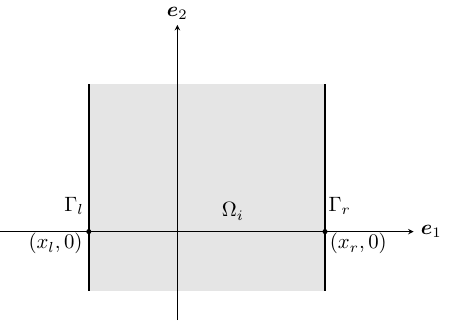}
\end{center}
\caption{\label{fig:inf-strip} The figure shows the computational
domain, $\Omega_i$, which is an infinite strip with boundaries parallel to the axis
$\vv{e}_2$. This domain can be replaced by a rectangular
domain if we assume periodic boundary condition along the unbounded direction.}
\end{figure}

\subsection{Infinite strip}\label{sec:DSB}
In this section, we restrict ourself to the case of an infinite strip with the 
boundary parallel to one of the coordinate axes, or, rectangular domain with a periodic 
boundary condition along one of the coordinate axes. Let the coordinate axes be
labelled as $\vv{e}_1$ and $\vv{e}_2$. For the infinite strip, say, with a 
boundary parallel to the axis $\vv{e}_2$, we assume that $\supp_{x_1}u_0(\vv{x})$
is bounded. The derivation of the TBCs for the infinite straight boundary is 
particularly simple and can be obtained using Laplace transform in time and a 
Fourier transform in space. Let us consider the IVP in~\eqref{eq:2D-SE} where 
the domain is defined by the infinite strip between $x_1=x_l$ and $x_1=x_r$ (see
Fig.~\ref{fig:inf-strip}). For the TBC on the right boundary, one must consider 
the IBVP on the right exterior domain 
$\Omega_r=(x_r,\infty)\times\field{R}$. Let the covariables of $(x_1,x_2)$ be denoted by 
$(\zeta_1,\zeta_2)$, respectively. We introduce the notation 
$\fourier_{x_1}f(x_1,x_2,t)=\fourier_{x_1}[f](\zeta_1,x_2,t)$ for one dimensional Fourier 
transform with respect to $x_1$ (similarly 
$\fourier_{x_2}f(x_1,x_2,t)=\fourier_{x_2}[{f}](x_1,\zeta_2,t)$ for Fourier transform with
respect to $x_2$). For denoting the Laplace transform of a function of $t$, 
we use $\laplace_t f(t)=\laplace_t[f](s)=F(s)$.

Let us denote the Fourier transform with respect $x_2$ of $w(x_1,x_2,t)$ by
$\widetilde{w}(x_1,\zeta_2,t)$ and Laplace transform
with respect to $t$ of $\widetilde{w}(x_1,x_2,t)$ by $\widetilde{W}(x_1,\zeta_2, s)$, i.e.,
\begin{equation}
\begin{split}
&\widetilde{w}(x_1,\zeta_2, s) = \fourier_{x_2}w(x_1,x_2,t),\\
&\widetilde{W}(x_1,\zeta_2, s) = \laplace_t\widetilde{w}(x_1,\zeta_2,t).
\end{split}
\end{equation}
For the case of compactly supported initial data, we have
\begin{equation}
(\partial_{x_1}^2+\alpha^2)\widetilde{W}(x_1,\zeta_2,s)=0,\quad x_1\in(x_r,\infty),
\end{equation}
where $\alpha=\sqrt{is-\zeta^2_2}$ such that $\Im(\alpha)>0$. The solution
can be worked out as follows: observing
\begin{equation*}
\begin{split}
&\partial_{x_1}\widetilde{W}(x_1,\zeta_2,s)
=i\alpha\widetilde{W}(x_1,\zeta_2,s),\\
&\laplace^{-1}[\partial_{x_1}\widetilde{W}(x_1,\zeta_2,s)]
=\laplace^{-1}[i\alpha^{-1}]\star\laplace^{-1}[\alpha^2\widetilde{W}(x_1,\zeta_2,s)],
\end{split}
\end{equation*}
where `$\star$' represents the convolution operation, we have
\begin{equation}\label{eq:DtN-inf-strip}
\partial_{x_1}\widetilde{w}(x_1,\zeta_2,t)=e^{i\pi/4}e^{-i\zeta_2^2t}\partial_t^{-1/2}e^{i\zeta_2^2t}\\
\left[
i\partial_t\widetilde{w}(x_1,\zeta_2,t)
+\widetilde{(\partial^2_{x_2}w)}(x_1,\zeta_2,t)\right].
\end{equation}
It is also easy to verify
\begin{equation}
\partial_t\left[e^{-i\zeta_2^2t}\partial_t^{-1/2}e^{i\zeta_2^2t}\widetilde{w}(x_1,\zeta_2,t)\right]\\
=e^{-i\zeta_2^2t}\partial_t^{-1/2}e^{i\zeta_2^2t}\partial_t\widetilde{w}(x_1,\zeta_2,t).
\end{equation}
Let
\begin{equation}
\mathcal{G}(x_2,t)=\frac{e^{-i\pi/4}}{\sqrt{4\pi t}}e^{i\frac{x_2^2}{4t}},
\quad t\in\field{R}_+,
\end{equation}
so that its Laplace transform reads as
$\widetilde{\mathcal{G}}(\zeta_2,t)=e^{-i\zeta_2^2t}$. Now, taking the inverse Fourier 
transform in~\eqref{eq:DtN-inf-strip}, we obtain the Dirichlet--to--Neumann map as~\cite{Menza1997}
\begin{widetext}
\begin{equation}\label{eq:IS-TBC}
\begin{split}
\partial_{x_1}{w}(\vv{x},t)
&=\frac{e^{i\pi/4}}{\sqrt{\pi}}\int_0^t\int_{\field{R}}
\left[i\partial_{\tau}{w}(x_1,x'_2,\tau)+\partial^2_{x'_2}w(x_1,x'_2,\tau)\right]
\frac{\mathcal{G}(x_2-x'_2,t-\tau)}{\sqrt{t-\tau}}dx_2'd\tau\\
&=-(\partial_{t}-i\partial^2_{x_2})\frac{e^{-i\pi/4}}{\sqrt{\pi}}\int_0^t
\int_{\field{R}}w(x_1,x'_2,\tau)\frac{\mathcal{G}(x_2-x'_2,t-\tau)}{\sqrt{t-\tau}}dx_2'd\tau,
\end{split}
\end{equation}
\end{widetext}
This map can be expressed compactly if we employ the notation
\begin{equation}\label{eq:Op-2D-TBC}
(\partial_t-i\partial_{x_2}^2)^{-1/2}f({x}_2,t)\\
=\frac{1}{\sqrt{\pi}}\int_0^t\int_{\field{R}}f(x_1,x'_2,\tau)
\frac{\mathcal{G}(x_2-x'_2,t-\tau)}{\sqrt{t-\tau}}dx_2'd\tau,
\end{equation}
where $f\in\fs{C}^{\infty}(\field{R}\times\field{R}_+)$ belongs to the Schwartz class with 
respect to $x_2$. On account of the singular nature of the
symbol, it is not a pseudo-differential operator. However, away from the points satisfying 
$\xi+\zeta_2^2=0$, it can be 
microlocally regarded as a pseudo-differential operator with a symbol $(i\xi+i\zeta_2^2)^{-1/2}$, with the 
branch cut defined by $-\pi\leq\arg{(i\xi+i\zeta_2^2)}<\pi$ where $(\zeta_2,\xi)$ are the covariables 
of $(x_2,t)$. Further, if we take the operator
$(\partial_t-i\partial_{x_2}^2)^{1/2}$ to be defined by
\begin{equation}
(\partial_t-i\partial_{x_2}^2)^{1/2}f=(\partial_t-i\partial_{x_2}^2)[(\partial_t-i\partial_{x_2}^2)^{-1/2}f],
\end{equation}
then
\begin{equation}
\partial_{x_1}{w}(\vv{x},t)+e^{-i\pi/4}(\partial_{t}-i\partial^2_{x_2})^{1/2}w(\vv{x},t)=0,
\end{equation}
for $(\vv{x},t)\in\Omega_r\times\field{R}_+$.

Next, we would like to obtain a more local approximation of this boundary condition valid for small times. 
To this end, let us consider
\begin{equation}
\laplace^{-1}[i\alpha^{-1}]=\frac{1}{2\pi}\int(is-\zeta^2_2)^{-{1}/{2}}e^{st}ds.
\end{equation}
Setting $\xi=st$, we have
\begin{equation*}
\begin{split}
\laplace^{-1}\left[\frac{i}{\alpha}\right]
%&=\frac{t^{-\frac{1}{2}}}{2\pi}\int_{a+i\field{R}}\frac{e^{\xi}}{\sqrt{i\xi-\zeta^2_2t}}d\xi\quad(a>0)\\
&=\frac{t^{-\frac{1}{2}}}{2\pi}\int_{a+i\field{R}}
\left(1-\frac{\zeta^2_2t}{i\xi}\right)^{-\frac{1}{2}}\frac{e^{\xi}d\xi}{\sqrt{i\xi}}\quad(a>0)\\
&\sim\frac{t^{-\frac{1}{2}}}{2\pi}\int_{a+i\field{R}}
\left(\frac{1}{(i\xi)^{\frac{1}{2}}}
+\frac{\zeta^2_2t}{2(i\xi)^{\frac{3}{2}}}
+\frac{3\zeta^4_2t^2}{8(i\xi)^{\frac{5}{2}}}
+\ldots\right)e^{\xi}d\xi\\
&\sim \frac{e^{i\pi/4}}{\Gamma(\tfrac{1}{2})}t^{-\frac{1}{2}}
+\frac{e^{-i\pi/4}}{2\Gamma(\tfrac{3}{2})}t^{\frac{1}{2}}\zeta_2^2
-\frac{3e^{i\pi/4}}{8\Gamma(\tfrac{5}{2})}t^{\frac{3}{2}}\zeta_2^4
+\ldots
\end{split}
\end{equation*}
Taking the inverse Fourier transform, we obtain the asymptotic expansion as
\begin{equation}\label{eq:ABC-2D-HF}
\partial_{x_1}w+e^{-i\pi/4}\partial_t^{1/2}w
-e^{i\pi/4}\frac{1}{2}\partial_{x_2}^2\partial_t^{-1/2}w=0\mod{(\partial_t^{-3/2})}.
\end{equation}

For the periodic case, we may take the computational domain to be 
$\Omega_i =(x_l,x_r)\times(0,2\pi)$ so that
\begin{equation}
\begin{split}
&w(x_1,x_2,t)=\sum_{m\in\field{Z}}\widetilde{w}_m(x_1,t)e^{im x_2},\\
&W(x_1,x_2,s)=\sum_{m\in\field{Z}}\widetilde{W}_m(x_1,s)e^{im x_2},
\end{split}
\end{equation}
with $\widetilde{W}_m(x_1,s)=\laplace_t\widetilde{w}_m(x_1,t)$. We also recall
$x_1\in(x_r,\infty)$ whereby a limiting procedure can be used to obtain the BCs
at $x_1=x_r$. For the sake of
simplicity, we demonstrate the procedure for the $m$-th Fourier component. The
complete result, then, follows by superposing all the components. Putting
$\alpha_m=\sqrt{is-m^2}$, the following results can be obtained in the same
manner as before:
\begin{equation}
\partial_{x_1}\widetilde{w}_m(x_1,t)=e^{i\pi/4}
e^{-i m^2t}\partial_t^{-1/2}e^{i m^2t}
\left[i\partial_t\widetilde{w}_m(x_1,t)-m^2\widetilde{w}_m(x_1,t)\right].
\end{equation}
Following the steps in the previous case, we have
\begin{equation*}
\partial_t\left[e^{-im^2t}\partial_t^{-1/2}e^{im^2t}\widetilde{w}_m(x_1,t)\right]
=e^{-im^2t}\partial_t^{-1/2}e^{im^2t}\partial_t\widetilde{w}_m(x_1,t),
\end{equation*}
and
\begin{equation}\label{eq:ISP-Op2}
e^{-i m^2t}\partial_t^{-1/2}e^{i m^2t}
\left[i\partial_t\widetilde{w}_m(x_1,t)-m^2\widetilde{w}_m(x_1,t)\right]\\
=ie^{-im^2t}\partial_t^{1/2}e^{im^2t}\widetilde{w}_m(x_1,t).
\end{equation}
Introducing
\begin{equation}\label{eq:kernel-periodic}
\mathcal{G}(x_2,t)=\sum_{m\in\field{Z}}e^{im x_2-im^2t},
\end{equation}
which is defined only in a distributional sense and using the relations above, 
we obtain an expression similar to~\eqref{eq:IS-TBC} as follows:
\begin{widetext}
\begin{equation}\label{eq:IPS-TBC}% Infinite periodic strip
\begin{split}
\partial_{x_1}{w}(\vv{x},t)
&=\frac{e^{i\pi/4}}{\sqrt{\pi}}\int_0^t\int_{0}^{2\pi}
\left[i\partial_{\tau}{w}(x_1,x'_2,\tau)+\partial^2_{x'_2}w(x_1,x'_2,\tau)\right]
\frac{\mathcal{G}(x_2-x'_2,t-\tau)}{\sqrt{t-\tau}}dx_2'd\tau\\
&=-(\partial_{t}-i\partial^2_{x_2})\frac{e^{-i\pi/4}}{\sqrt{\pi}}
\int_0^t\int_{0}^{2\pi}w(x_1,x'_2,\tau)\frac{\mathcal{G}(x_2-x'_2,t-\tau)}{\sqrt{t-\tau}}dx_2'd\tau.
\end{split}
\end{equation}
\end{widetext}

For periodic functions, the operator defined in~\eqref{eq:Op-2D-TBC} becomes 
\begin{equation}\label{eq:Op-ISP-TBC}
(\partial_t-i\partial_{x_2}^2)^{-1/2}f({x}_2,t)\\
=\frac{1}{\sqrt{\pi}}\int_0^t\int_{0}^{2\pi}f(x_1,x'_2,\tau)
\frac{\mathcal{G}(x_2-x'_2,t-\tau)}{\sqrt{t-\tau}}dx_2'd\tau,
\end{equation}
where the kernel is defined by~\eqref{eq:kernel-periodic}. We now turn our
attention to obtaining a form of the operator
$(\partial_t-i\partial_{x_2}^2)^{1/2}$ which can be numerically implemented.
Using the relation in~\eqref{eq:ISP-Op2}, we may write
\begin{equation*}
\begin{split}
\partial_{x_1}{w}(\vv{x},t)
&=-{e^{-i\pi/4}}\sum_{m\in\field{Z}}e^{-im^2t}\partial^{1/2}_t
\left[e^{im^2t}\widetilde{w}_m(x_1,t)e^{imx_2}\right],\\
&=-{e^{-i\pi/4}}\partial^{1/2}_{t'}\sum_{m\in\field{Z}}
\left[e^{-im^2(t-t')}\widetilde{w}_m(x_1,t')e^{imx_2}\right]_{t'=t}.
\end{split}
\end{equation*}
Introducing the auxiliary function $\varphi(x_1,x_2,t,t')$ such that
\begin{equation}
\begin{split}
&\varphi(x_1,x_2,t,t')
=\sum_{m\in\field{Z}}\left[e^{-im^2(t-t')}\widetilde{w}_m(x_1,t')e^{imx_2}\right],\\
&\partial_{x_1}{w}(\vv{x},t)=\left.-{e^{-i\pi/4}}\partial^{1/2}_{t'}\varphi(x_1,x_2,t,t')\right|_{t'=t}.
\end{split}
\end{equation}
%%%%%%%%%%%%%%%%%%%%%%%%%%
\begin{figure}[!htbp]
\begin{center}
\def\myscale{1}
\includegraphics[scale=\myscale]{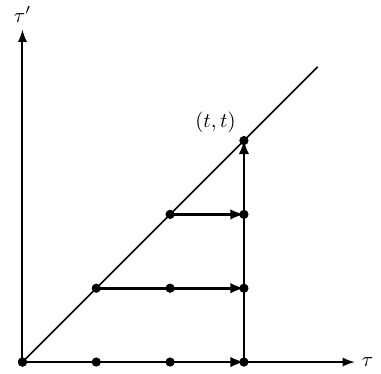}
\end{center}
\caption{\label{fig:IVP-auxi-periodic} In this figure, a schematic is shown 
illustrating how the auxiliary equation~\eqref{eq:FSE-auxi-ivp} will be solved 
in order to provide the history of the field needed in the TBCs. The field is known 
along the diagonal which serves as the initial conditions to obtain the values of the field needed
in the TBCs (arrow in the line depicts the evolution direction in time).}
\end{figure}
In order to determine all the values of the function needed to compute the non-local
fractional derivative, consider the IVP given by
\begin{equation}\label{eq:FSE-auxi-ivp}
\left\{\begin{aligned}
&[i\partial_{\tau}+\partial^2_{x_2}]\varphi(x_1,x_2,\tau,\tau')=0,
\quad(\tau,x_2)\in(\tau',t]\times(0,2\pi),\\
&\varphi(x_1,0,\tau',\tau')=\varphi(x_1,2\pi,\tau',\tau'),\quad\tau\in(\tau',t],\\
&\varphi(x_1,x_2,\tau',\tau')=w(x_1,x_2,\tau'),\quad x_2\in(0,2\pi).
\end{aligned}\right.
\end{equation}
The solution of this IVP must be obtained for all $\tau'\in[0,t]$. This process is
schematically depicted in Fig.~\ref{fig:IVP-auxi-periodic} where we note that the history of the
field is needed along the vertical line up to the diagonal in the $(\tau,\tau')$-plane.
\begin{rmk}\label{rmk:periodic}
Such a procedure can be used to compute the action of any operator of the form
$(\partial_t-i\partial^2_{x})^{-n/2},\,\,n=1,2,\ldots,$ on any function periodic in $x\in\field{R}$, say,
$f(x,t)$. Introducing an auxiliary function $\varphi(x,t,t')$ such that
\begin{equation}
\begin{split}
&\varphi(x,t,t')
=\sum_{m\in\field{Z}}\left[e^{-im^2(t-t')}\widetilde{f}_m(t')e^{imx}\right]_{t'=t},\\
&(\partial_t-i\partial^2_{x})^{-n/2}f(x,t)=\partial^{-n/2}_{t'}\varphi(x,t,t')|_{t'=t}.
\end{split}
\end{equation}
The associated IVP is given by
\begin{equation}
\left\{\begin{aligned}
&[i\partial_{\tau}+\partial^2_{x}]\varphi(x,\tau,\tau')=0,\quad(\tau,x)\in(\tau',t]\times(0,2\pi),\\
&\varphi(0,\tau,\tau') = \varphi(2\pi,\tau,\tau'),\quad\tau\in(\tau',t],\\
&\varphi(x,\tau',\tau')=f(x,\tau'),\quad x\in(0,2\pi),
\end{aligned}\right.
\end{equation}
which needs to be solved for all $\tau'\in[0,t]$.
\end{rmk}

\subsubsection{Stability and uniqueness}
An equivalent formulation of the IVP~\eqref{eq:2D-SE} on the computational
domain $\Omega_i = (x_l,x_r)\times(0, 2\pi)$ with periodic boundary condition
along the axis $\vv{e}_2$ is given by
\begin{equation}\label{eq:2D-SE-Omegai}
\left\{\begin{aligned}
&i\partial_tu+\triangle u=0,\quad (\vv{x},t)\in\Omega_i\times\field{R}_+,\\
&u(\vv{x},0)=u_0(\vv{x})\in \fs{L}^2(\Omega_i),\quad\supp\, u_0\subset\Omega_i,\\
&u(x_1, x_2+2\pi,t)=u(x_1, x_2,t),\quad t>0,\\
&\partial_{n}{u}+
e^{-i\pi/4}(\partial_{t}-i\partial^2_{x_2})^{1/2}u=0,
\quad\vv{x}\in\Gamma_l\cup\Gamma_r,\,t>0.
\end{aligned}\right.
\end{equation}
Setting $d\vbs{\varsigma}=\vv{e}_n|d\vv{x}|$ and assuming that the solution 
$u(\vv{x},t)$ exists, we have
\begin{equation}
\int_{\Omega_i}(\partial_t|u|^2)d^2\vv{x}
=2\Re\int_{\Gamma_l\cup\Gamma_r}(u^*i\nabla u)\cdot d\vbs{\varsigma},
\end{equation}
so that
\begin{equation*}
\|u(\cdot,T)\|^2_{\fs{L}^2(\Omega_i)}-\|u_0\|^2_{\fs{L}^2(\Omega_i)}
=\Re\int_0^T dt\int_{\Gamma_l\cup\Gamma_r}(u^*i\nabla u)\cdot
d\vbs{\varsigma},
\end{equation*}
where  for $\vv{x}\in\Gamma$.
Further, noting $\tilde{u}_m(x_1,\cdot)\in \fs{H}^{1/4}([0, T])$ (denotes the Sobolev 
space $\fs{W}^{{1}/{4},2}([0, T])$), we have
\begin{equation*}
\begin{split}
\mathcal{I}_R
&=-2\Re\int_{\field{R}_+}
dt\int_{\Gamma_r}\left(u^*e^{i\pi/4}\sqrt{\partial_t-i\partial^2_{x_2}}u\right)dx_2,\\
&=-4\pi\Re\sum_{m\in\field{Z}}\int_{\field{R}_+}dt
\overline{\left(\tilde{u}_m(x_r,t)e^{-im^2t}\right)}
\left(\partial_t^{1/2}\tilde{u}_m(x_r,t)e^{-im^2t}\right),\\
&=-2\sum_{m\in\field{Z}}\int_{\field{R}_-}d\xi|\xi|^{1/2}
\left|\fourier_t\left[\tilde{u}_m(x_r,t)e^{-im^2t}\right](x_r,\xi)\right|^2\leq0.
\end{split}
\end{equation*}
Similarly, it can be shown that
\begin{equation*}
\mathcal{I}_L
=-2\Re\int_{\field{R}_+}
dt\int_{\Gamma_l}\left(u^*e^{i\pi/4}\sqrt{\partial_t-i\partial^2_{x_2}}u\right)dx_2\leq0.
\end{equation*}
Therefore, we have 
\begin{equation*}
\|u(\cdot,T)\|_{\fs{L}^2(\Omega_i)}\leq\|u_0\|_{\fs{L}^2(\Omega_i)}.
\end{equation*}
This result also guarantees the uniqueness of the solution.  

%%%%%%%%%%%%%%%%%%%%%%%%%%%%%%%%%%%%%%%%%%%%%%%%%%%%%%%%%%%%%%%%%%%%%%%%%%%%%%%%%%%%%%%%%%%%%%%%%%
\subsection{Rectangular domains}\label{sec:rect-domain}
Let us consider a rectangular domain given by 
\begin{equation}
\Omega_i=\{(x_1,x_2):x_b<x_2<x_t, x_l<x_1<x_r\},
\end{equation}
and denote the boundaries as $\Gamma_{l,r}=\{(x_1,x_2)\in\partial\Omega_i: x_1=x_{l,r}\}$, 
respectively and $\Gamma_{b,t}=\{(x_1,x_2)\in\partial\Omega_i: x_1=x_{b,t}\}$, respectively.
Assuming $\supp\,u_0$ bounded in $\Omega_i$, the TBCs for the
infinite strip cannot be taken to be the transparent boundary operators at the
straight edges because the corresponding operator requires knowledge of the entire 
support of the field along the tangential direction at the boundary. This clearly cannot 
be achieved because after a certain time the field would have spread outside the
domain defined by any segment of the rectangular domain, say, $\Gamma_r$. This issue can be resolved in the
following way: Using the same notation as in the last section and observing that
\begin{figure}[!htbp]
\begin{center}
\def\myscale{1}
\includegraphics[scale=\myscale]{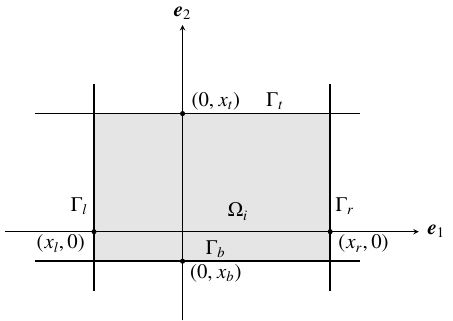}
\end{center}
\caption{\label{fig:rect-domain} The figure shows a rectangular domain
with boundary segments parallel to one of the axes.}
\end{figure}

\begin{figure*}[!hbt]
\begin{center}
\def\myscale{1}
\includegraphics[scale=\myscale]{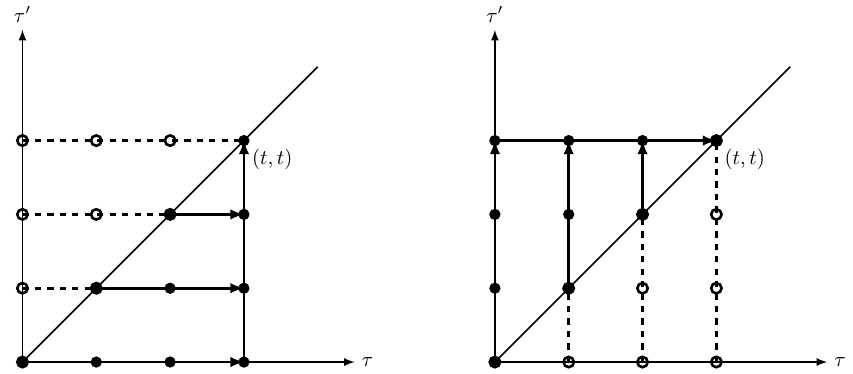}
\end{center}
\caption{\label{fig:IVP-auxi}A schematic depiction of the evolution of the 
auxiliary field $\varphi(x_1,x_2,\tau,\tau')$ in the 
$(\tau,\tau')$-plane is provided in this figure where the plot on the left corresponds 
$\vv{x}\in\Gamma_r\cup\Gamma_l$ and the plot on the right corresponds 
$\vv{x}\in\Gamma_t\cup\Gamma_b$. The filled circles denote the evolution of any arbitrary
point that belongs to the boundary. The empty circles are relevant only when the
point under consideration is a corner point. The evolution over the interior 
of the domain is carried either above or below the diagonal starting from the diagonal 
values (where direction of evolution is depicted by the arrow head). In
contrast, the corner points 
are evolved on either side of the diagonal. Note that the vertical/horizontal lines where
the arrows end corresponds to the history of the auxiliary field needed for
TBCs in the current time ($t$). The TBCs for auxiliary field require the history 
of the auxiliary field at the corner points, depicted by broken lines; these values
are taken from the adjacent segment of the boundary.}
\end{figure*}

\begin{equation*}
\begin{split}
&\partial_{x_1}\widetilde{w}(x_1,\zeta_2,t)
=e^{i\pi/4}e^{-i\zeta_2^2t}\partial_t^{-1/2}e^{i\zeta_2^2t}
\left[i\partial_t\widetilde{w}(x_1,\zeta_2,t)
+\widetilde{(\partial^2_{x_2}w)}(x_1,\zeta_2,t)\right],\\
&\partial_{x_1}\widetilde{w}(x_1,\zeta_2,t)
=-e^{-i\pi/4}\partial_{t'}^{1/2}e^{-i\zeta_2^2(t-t')}
\left.\widetilde{w}(x_1,\zeta_2,t')\right|_{t'=t},
\end{split}
\end{equation*}
and introducing the auxiliary function
\begin{equation}
\fourier_{x_2}{\varphi}(x_1,x_2,t,t')=e^{-i\zeta_2^2(t-t')}
\widetilde{w}(x_1,\zeta_2,t')
\end{equation}
so that
\begin{equation}
\partial_{x_1}{w}(x_1,x_2,t)
=-e^{-i\pi/4}\partial_{t'}^{1/2}\left.{\varphi}(x_1,x_2,t,t')
\right|_{t'=t}.
\end{equation}
It is easy to verify that $\varphi$ satisfies the IVP given by
\begin{equation}
\begin{split}
&[i\partial_{\tau}+\partial^2_{x_2}]\varphi(x_1,x_2,\tau,\tau')=0,\quad\in(\tau',t],\\
&\varphi(x_1,x_2,\tau',\tau')=w(x_1,x_2,\tau'),
\end{split}
\end{equation}
which needs to be solved for $\tau'\in[0,t]$ over $\Gamma_r$. The boundary conditions at the 
endpoints of the segment $\Gamma_r$ are not
known because the original problem is defined on the infinite domain $x_2\in \field{R}$.
However, it can be shown that the restriction of the auxiliary function
$\varphi(x_1,x_2,\tau,\tau')$ to $x_1=x_r$ at $\tau=0$ is compactly 
supported (with respect to $x_2$). This would allow one to impose the transparent 
boundary conditions at the endpoints of
$\Gamma_r$. To this end, let us consider the IVP defined by~\eqref{eq:2D-SE}, it can 
be solved using Fourier transform in $(x_1,x_2)$. Putting 
$\tilde{u}_0(\vbs{\zeta})=\fourier_{(x_1,x_2)}u_0(\vv{x})$, we have
\begin{equation*}
\begin{split}
&\fourier_{(x_1,x_2)}u(\vv{x},t')=e^{-i(\zeta_1^2+\zeta_2^2)\tau'}
\tilde{u}_0(\vbs{\zeta}),\\
&\varphi(x_1,x_2,\tau,\tau')=\frac{1}{(2\pi)^2}\int_{\field{R}^2}
e^{i\vbs{\zeta}\cdot\vv{x}-i\zeta_1^2t'-i\zeta_2^2\tau}
\tilde{u}_0(\vbs{\zeta})d^2\vbs{\zeta}.
\end{split}
\end{equation*}
Therefore, $\supp_{x_2}\varphi(x_r,x_2,0,\tau')\subset[x_b,x_t]$ and 
\begin{equation}
\begin{split}
&[i\partial_{\tau'}+\partial^2_{x_1}]\varphi(x_1,x_2,\tau,\tau')=0,
\quad\tau'\in(\tau,t],\\
&\varphi(x_1,x_2,\tau,\tau)=w(x_1,x_2,\tau),\quad\tau\in[0,t].
\end{split}
\end{equation}
Hence, the boundary condition at $(x_1,x_2)\in\partial\Gamma_r$ is given by
\begin{equation}
\partial_{x_2}\varphi(x_1,x_2,\tau,\tau')
\pm e^{-i\pi/4}\partial^{1/2}_{\tau}\varphi(x_1,x_2,\tau,\tau')=0,
\end{equation}
where the sign is determined by $x_2\in\{x_t, x_b\}$, respectively. This requires 
the knowledge of $\varphi(x_1,x_2,\tau,\tau')$ at $(x_1,x_2)\in\partial\Gamma_r$ 
and $0\leq \tau\leq \tau'$. The algorithm 
can be explained by means of the Fig.~\ref{fig:IVP-auxi}. There are two IVPs for the 
auxiliary field $\varphi(x_1,x_2,\tau,\tau')$ each of which evolve the field either above or below the
diagonal in the $(\tau,\tau')$-plane starting from the values at the diagonal in their
respective domains. The filled circles denote the evolution of any arbitrary
point of the boundary and the arrows denote the direction of evolution. All diagonal 
points are evolved along $\tau$ or $\tau'$-axis in order to provide the history needed 
for TBCs in the current time step (represented by horizontal or vertical line in 
$(\tau,\tau')$-plane). The empty circles are only relevant for the corner
points. The values of the auxiliary field at the corners are needed 
for the TBCs satisfied by the auxiliary fields; this relationship is depicted by the broken
lines. Note that these values at the corners can be taken from the adjacent segment of 
the boundary where it is already being computed. 

\begin{rmk}
From the discussion above, we obtain the following useful definition of the operator
$(\partial_t-i\partial^2_{x})^{1/2}$:
\begin{equation}
(\partial_t-i\partial_{x}^2)^{1/2}f(x,t)\\=\frac{1}{2\pi}\iint d\zeta dx'
e^{-i\zeta(x-x')}[\partial_{t'}^{1/2}e^{-i\zeta^2(t-t')}f(x',t')]_{t'=t},
\end{equation}
for $(x,t)\in\field{R}\times\field{R}_+$. This definition makes it explicit that one 
has to consider the function
$f(x,t)$ over its entire support with respect to $x\in\field{R}$ in order to
compute the expression; however, in special cases this can be avoided.
Similarly, a formal definition of the operator $(\partial_t-i\partial^2_{x})^{-m/2},
(x,t)\in\field{R}\times \field{R}_+,\,\,m>0,$ for any arbitrary function $f(x,t)$ with sufficient
smoothness property can be given. This operator can be defined as
\begin{equation}
(\partial_t-i\partial_{x}^2)^{-m/2}f(x,t)\\=
\frac{1}{2\pi}\iint d\zeta dx'e^{-i\zeta(x-x')}[\partial_{t'}^{-m/2}e^{-i\zeta^2(t-t')}f(x',t')]_{t'=t}.
\end{equation}
The integration with respect to $\zeta$ can be easily performed. Defining the convolution 
kernel $\mathcal{K}(x,t)$ by
\begin{equation}
\mathcal{K}(x,t)=
\begin{cases}
\frac{e^{-i\pi/4}}{2\sqrt{\pi}\Gamma(m/2)}t^{-\frac{m+3}{2}}
\exp\left[i\frac{x^2}{4t}\right],&t>0,\\
0,&t<0,
\end{cases}
\end{equation}
we obtain 
\begin{equation}
(\partial_t-i\partial_{x}^2)^{-m/2}f(x,t)
=\iint dx'dt'\mathcal{K}(x-x',t-t')f(x',t').
\end{equation}
From this definition, it is clear that the operator cannot be defined on a
compact domain with respect to $x$ for arbitrary functions. Let us introduce
$\varphi(x,t,t')$ defined by
\begin{equation}
\varphi(x,t,t')=\frac{1}{2\pi}\iint d\zeta dx'
e^{-i\zeta(x-x')}e^{-i\zeta^2(t-t')}f(x',t'),
\end{equation}
so that 
\begin{equation}
(\partial_t-i\partial_{x}^2)^{-m/2}f(x,t)=\partial^{-m/2}_{t'}\varphi(x,t,t')|_{t'=t},
\end{equation}
where the auxiliary field satisfies the IVP given by
\begin{equation}\label{eq:associated-IVP}
\begin{split}
&i\partial_{\tau}\varphi(x,\tau,\tau')+\partial_x^2\varphi(x,\tau,\tau')=0,\quad\tau\in(\tau',t],\\
&\varphi(x,\tau',\tau') = f(x,\tau').
\end{split}
\end{equation}
Again, the solution of the IVP must be obtained for all $\tau'\in[0,t]$. If
$f(x,\tau)$ has a bounded support, $\Omega$, with respect to $x$ for all
$\tau\in[0,t]$, then the IVP above can be solved under the Dirichlet boundary
condition on $\partial\Omega$ (which is not a significantly different situation
that the periodic case discussed in Remark~\ref{rmk:periodic})
\end{rmk}

\subsubsection{High-frequency approximation\label{sec:HF-rect}}
The high-frequency approximation affords the possibility of simplifying the TBC
by making them ``local'' in terms of the spatial variable. The asymptotic
expansion worked out in~\eqref{eq:ABC-2D-HF} can be carried out for each of the edges 
of the rectangular domain to obtain the following ABCs:
\begin{equation}\label{eq:ABC-2D-HF-rect}
\begin{split}
&\partial_{n}u+e^{-i\pi/4}\partial_t^{1/2}u
-e^{i\pi/4}\frac{1}{2}\partial_{x_2}^2\partial_t^{-1/2}u=0,
\quad\vv{x}\in\Gamma_r\cup\Gamma_l,\\
&\partial_{n}u+e^{-i\pi/4}\partial_t^{1/2}u
-e^{i\pi/4}\frac{1}{2}\partial_{x_1}^2\partial_t^{-1/2}u=0,
\quad\vv{x}\in\Gamma_b\cup\Gamma_t.
\end{split}
\end{equation}
These boundary conditions become problematic at the corners of the rectangular
domain. This aspect can be illustrated by the considering the weak formulation
of the original IVP as follows: Consider a test function 
$\psi(\vv{x})\in\fs{W}^{1,2}(\Omega_i)$ (Sobolev space); taking the inner product with the 
equation~\eqref{eq:2D-SE}, we have
\begin{equation*}
\int_{\Omega_i}(i\partial_tu+\nabla^2u)\psi d^2\vv{x}\\
=i\partial_t\int_{\Omega_i}u\psi d^2\vv{x}
-\int_{\Omega_i}(\nabla u)\cdot(\nabla\psi)d^2\vv{x}
+\int_{\partial\Omega_i}\psi(\nabla u)\cdot d\vbs{\varsigma}.
\end{equation*}
Let us consider the top and right boundaries. The boundary integrals are given by
% \begin{equation*}
% \mathcal{I}_R+\mathcal{I}_T
% =\int_{\Gamma_r}\psi\partial_{{x_1}}u dx_2 +\int_{\Gamma_t}\psi\partial_{{x_2}}u dx_1,
% \end{equation*}
%which on simplification reads as
%\begin{widetext}
\begin{equation*}
\begin{split}
%&\mathcal{I}_R+\mathcal{I}_T\\
&\int_{\Gamma_r}\psi\partial_{{x_1}}u dx_2 
+\int_{\Gamma_t}\psi\partial_{{x_2}}u dx_1\\
&=-e^{-i\pi/4}\int_{\Gamma_r\cup\Gamma_l}\psi\partial^{1/2}_tu
+\frac{1}{2}e^{i\pi/4}\left[\int_{\Gamma_r}\psi\partial^2_{{x_2}}\partial^{-1/2}_tu dx_2
+\int_{\Gamma_t}\psi\partial^2_{{x_1}}\partial^{-1/2}_tu dx_1\right]\\
&=-e^{-i\pi/4}\int_{\Gamma_r\cup\Gamma_l}\psi\partial^{1/2}_tu
+\frac{1}{2}e^{i\pi/4}\left[
\left.\psi\partial_{x_2}\partial^{-1/2}_tu\right|_{x_2=x_b}^{x_t}+
\left.\psi\partial_{x_1}\partial^{-1/2}_tu\right|_{x_1=x_l}^{x_r}\right]\\
&\qquad-\frac{1}{2}e^{i\pi/4}\left[\int_{\Gamma_r}(\partial_{x_2}\psi)(\partial_{{x_2}}\partial^{-1/2}_tu) dx_2
+\int_{\Gamma_t}(\partial_{x_1}\psi)(\partial_{{x_1}}\partial^{-1/2}_tu)dx_1\right].
\end{split}
\end{equation*}
%\end{widetext}
Consider the terms which correspond to the top-right corner in the above
equation:
\begin{equation}
\left(\partial_{x_2}\partial^{-1/2}_tu
+\partial_{x_1}\partial^{-1/2}_tu\right)_{\Gamma_r\cap\Gamma_t}
=\partial_t^{-1/2}\left(\partial_{x_2}u+\partial_{x_1}u\right)_{\Gamma_r\cap\Gamma_t}.
\end{equation}
They are problematic on account of the fact that the BCs in the current 
form cannot be used to evaluate them. In order to evaluate these terms, we carry out the
fractional integration, $\partial^{-1/2}_t$, of the evolution
equation in~\eqref{eq:2D-SE} so that
\begin{equation}
i\partial_t^{1/2}u+(\partial^2_{x_1}+\partial^2_{x_2})\partial_t^{-1/2}u=0,
\,\,(x_1,x_2)\in\Gamma_r\cap\Gamma_t.
\end{equation}
Here, the fact that the field is zero at the corner at $t=0$ is explicitly used
to arrive at the fractional derivative. Using BCs in~\eqref{eq:ABC-2D-HF-rect} and
the last equation, we obtained the following corner condition:
\begin{equation}
\partial_{x_1}u+\partial_{x_2}u+\frac{3}{2}e^{-i\pi/4}\partial_t^{1/2}u=0,
\,\,(x_1,x_2)\in\Gamma_r\cap\Gamma_t.
\end{equation}
A similar procedure can be used to the construct corner conditions for the other
corners of the rectangular domain:
\begin{equation}
\partial_{n}u|_{\Gamma_i}+\partial_{n}u|_{\Gamma_j}+\frac{3}{2}e^{-i\pi/4}\partial_t^{1/2}u=0,
\,\,(x_1,x_2)\in \Gamma_i\cap\Gamma_j,
\end{equation}
where $i\neq j$ and $i,j\in\{r,t,l,b\}.$

\subsubsection{Stability and uniqueness}
Let us write the equivalent formulation of the IVP~\eqref{eq:2D-SE} for a
rectangular domain $\Omega_i$ using the TBCs derived in the last section:
\begin{equation}\label{eq:2D-SE-Omegai-rect}
\left\{\begin{aligned}
&i\partial_tu+\triangle u=0,\quad(\vv{x},t)\in\Omega_i\times\field{R}_+,\\
&u(\vv{x},0)=u_0(\vv{x})\in \fs{L}^2(\Omega_i),\quad\supp\, u_0\subset\Omega_i,\\
&\partial_{n}{u}+e^{-i\pi/4}(\partial_{t}-i\partial^2_{x_2})^{1/2}u=0,
\quad\vv{x}\in\Gamma_l\cup\Gamma_r,\\
&\partial_{n}{u}+e^{-i\pi/4}(\partial_{t}-i\partial^2_{x_1})^{1/2}u=0,
\quad\vv{x}\in\Gamma_b\cup\Gamma_t,\,t>0.
\end{aligned}\right.
\end{equation}
Assuming that the solution $u(\vv{x},t)$ exits for $t\in[0,T]$, we have
\begin{equation}\label{eq:Energy-estimate}
\|u(\cdot,T)\|^2_{\fs{L}^2(\Omega_i)}-\|u_0\|^2_{\fs{L}^2(\Omega_i)}=\\
2\Re\int_0^T dt\left[\int_{\Gamma_l\cup\Gamma_r}(u^*i\nabla u)\cdot
d\vbs{\varsigma}+\int_{\Gamma_b\cup\Gamma_t}(u^*i\nabla u)\cdot
d\vbs{\varsigma}\right].
\end{equation}
In the above equation, the fields in the boundary integral can be extended to
$\field{R}_+$ without changing the value of the integral by setting them
identically to zero outside the interval $[0,T]$. For the right boundary, we have
\begin{equation}
\mathcal{I}_R
=-2\Re\int_0^Tdt
\int_{\Gamma_r}\left(u^*e^{i\pi/4}\sqrt{\partial_t-i\partial^2_{x_2}}u\right)dx_2.
\end{equation}
Again, without changing the value of the integral with respect to $x_2$, the fields 
can be extended to whole line by setting them identically to zero outside 
$\Gamma_r$. We these considerations in mind, one can write
\begin{equation*}
\begin{split}
\mathcal{I}_R
&=-\frac{1}{\pi}\Re\int_{\field{R}_+}dt\int_{\field{R}}d\zeta_2
\overline{\left(\tilde{u}(x_r,\zeta_2,t)e^{-i\zeta_2^2t}\right)}
\left(\partial_t^{\frac{1}{2}}\tilde{u}(x_r,\zeta_2,t)e^{-i\zeta_2^2t}\right),\\
&=-\frac{1}{2\pi^2}\int_{\field{R}_-}d\xi\int_{\field{R}}d\zeta_2
|\xi|^{\frac{1}{2}}
\left|\fourier_t\left[\tilde{u}(x_r,\zeta_2,t)e^{-i\zeta_2^2t}\right](x_r,\zeta_2,\xi)\right|^2
\end{split}
\end{equation*}
so that $\mathcal{I}_R\leq0$.
Similarly, it can be shown that the other boundary integrals namely
$\mathcal{I}_{L}$, $\mathcal{I}_{B}$ and $\mathcal{I}_{T}$ corresponding to the left,
bottom and top boundary, respectively, also satisfy the same inequality as
that of $\mathcal{I}_{R}$. Therefore, we have 
\begin{equation*}
\|u(\cdot,T)\|_{\fs{L}^2(\Omega_i)}\leq\|u_0\|_{\fs{L}^2(\Omega_i)}.
\end{equation*}
Since $T\in\field{R}_+$ is arbitrary, one replace $T$ with $t$ in the above
inequality. This result also guarantees the uniqueness of the solution.  

For the high-frequency approximation, we have the following equivalent
formulation on $\Omega_i$: Putting $\Gamma=\partial\Omega_i$ and $\Gamma_C$, 
the set of four corner points, denote the Laplace-Beltrami operator by
$\triangle_{\Gamma}$ so that
\begin{equation}\label{eq:2D-SE-Omegai-rect-HF}
\left\{\begin{aligned}
&i\partial_tu+\triangle u=0,\quad(\vv{x},t)\in\Omega_i\times\field{R}_+,\\
&u(\vv{x},0)=u_0(\vv{x})\in \fs{L}^2(\Omega_i),\quad\supp\,u_0\subset\Omega_i,\\
&\partial_{n}u+e^{-i\pi/4}\partial_t^{1/2}u
-e^{i\pi/4}\frac{1}{2}\triangle_{\Gamma}\partial_t^{-\frac{1}{2}}u=0,
\quad\vv{x}\in\Gamma\setminus\Gamma_C,\\
&\partial_{n}u|_{\Gamma_i}+\partial_{n}u|_{\Gamma_j}
+\frac{3}{2}e^{-i\pi/4}\partial_t^{\frac{1}{2}}u=0,
\,\vv{x}\in\Gamma_i\cap\Gamma_j,\,t>0,
\end{aligned}\right.
\end{equation}
where $i\neq j$, and $i,j\in\{r,t,l,b\}$. Taking equation~\eqref{eq:Energy-estimate} 
and following the standard approach, we have
\begin{equation*}
\begin{split}
\int_{\Gamma}(u^*i\nabla u)\cdot d\vbs{\varsigma}
&=-e^{i\pi/4}\int_{\Gamma}u^*\partial^{\frac{1}{2}}_{t}u|d\vv{x}|
-\frac{1}{2}e^{-i\pi/4}\int_{\Gamma\setminus\Gamma_C}
u^*(\triangle_{\Gamma}\partial_t^{-\frac{1}{2}}u)|d\vv{x}|\\
&=-e^{i\pi/4}\int_{\Gamma}u^*\partial^{\frac{1}{2}}_{t}u|d\vv{x}|
+\frac{1}{2}e^{-i\pi/4}\int_{\Gamma\setminus\Gamma_C}
(\partial_{\Gamma}u)^*(\partial_{\Gamma}\partial_t^{-\frac{1}{2}}u)|d\vv{x}|\\
&\qquad+\frac{3}{4}i\left(
|u(x_r, x_t)|^2-|u(x_l, x_t)|^2
+|u(x_l, x_b)|^2-|u(x_r, x_b)|^2\right).
\end{split}
\end{equation*}
Taking the real part and plugging the result back into equation~\eqref{eq:Energy-estimate} yields
\begin{widetext}
\begin{equation*}
\begin{split}
\|u(\cdot,T)\|^2_{\fs{L}^2(\Omega_i)}
-\|u_0\|^2_{\fs{L}^2(\Omega_i)}&
=2\Re\int_0^T dt\left[-e^{i\pi/4}\int_{\Gamma}u^*\partial^{1/2}_{t}u|d\vv{x}|
+\frac{1}{2}e^{-i\pi/4}\int_{\Gamma\setminus\Gamma_C}
(\partial_{\Gamma}u)^*(\partial_{\Gamma}\partial_t^{-1/2}u)|d\vv{x}|\right]\\
&=2\int_{\field{R}_-}d\xi\sum_{\Gamma_i}\int_{\field{R}}d\zeta
\left(-|\xi|^{1/2}+\frac{1}{2}|\xi|^{-1/2}|\zeta|^2\right)
|\fourier_t\fourier_{\varsigma}[u(\vv{x},t)]|^2,
\end{split}
\end{equation*}
\end{widetext}
where $\varsigma$ is taken as the tangential variable to the boundary so that
$\partial_{\Gamma}=\partial_{\varsigma}$. In the region defined by
$|\xi|\geq|\zeta|^2/2$, i.e., under the high-frequency approximation, we obtain
\begin{equation*}
\|u(\cdot,T)\|_{\fs{L}^2(\Omega_i)}\leq\|u_0\|_{\fs{L}^2(\Omega_i)}.
\end{equation*}

%==========================================================================
%==========================================================================
%==========================================================================
\section{General Schr\"odinger equation\label{sec:gse}}
Let us consider the IVP corresponding to the general Schr\"{o}dinger equation 
given by
\begin{equation}
\begin{split}\label{eq:2D-NL}
&i\partial_tu+\triangle u+\phi(\vv{x},t,|u|^2)u=0,
\quad(\vv{x},t)\in\field{R}^2\times\field{R}_+,\\
&u(\vv{x},0)=u_0(\vv{x}),\quad\vv{x}\in\field{R}^2.
\end{split}
\end{equation}
The potential function, $\phi$, is assumed to be real valued. Define
\begin{equation}\label{eq:Phi}
\Phi =\int_0^t\phi(\vv{x},t',|u(\vv{x},t')|^2)dt'.
\end{equation}
The linear Schr\"odinger equation with time-dependent potential and the
nonlinear case are treated in the same fashion in this section. 

\begin{figure}[!htbp]
\begin{center}
\def\myscale{1}
\includegraphics[scale=\myscale]{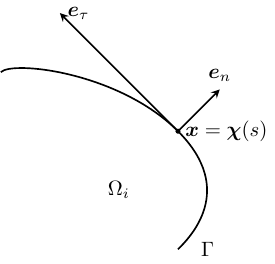}
\end{center}
\caption{\label{fig:computational-domain} The figure shows the parametrization of the boundary
$\Gamma$ of a convex domain $\Omega_i$.  Here, $\vv{e}_{n}$ and $\vv{e}_{\tau}$ denote the 
normal and the tangent vectors, respectively.}
\end{figure}
\subsection{Convex domain with smooth boundary}
The computational domain is taken to be a convex set $\Omega_i\subset\field{R}^2$ with a 
smooth boundary $\Gamma=\partial\Omega_i$. Let $\vbs{\chi}(s)$ be the parametrization of 
the curve $\Gamma$ where `$s$' is the length along the curve. Introducing the tangent vector 
$\vv{e}_{\tau}$, also a function of $s$ only, and taking into account the convexity of the domain 
we have
\begin{equation}
d\vv{x}=dr\vv{e}_n+(1+r\kappa)ds\vv{e}_{\tau},
\end{equation}
where $\kappa$ is the curvature of the boundary $\Gamma$ given by
\begin{equation}
\kappa(s)=\left|{\frac{d\vv{e}_{\tau}}{ds}}\right|
=\frac{|\det(\vbs{\chi}',\vbs{\chi}'')|}{|\vbs{\chi}'|^3},
\end{equation}
and $h = (1+r\kappa)$. The initial data is assumed to be compactly supported in the 
computational domain $\Omega_i$. Carrying out the gauge transformation 
$u=ve^{i\Phi}$ where $\Phi$ is defined by~\eqref{eq:Phi}, the evolution operator 
$L\equiv i\partial_t+\triangle+\phi$ in the curvilinear system is given by
\begin{equation}\label{eq:gauge-transform}
L(r,s,t,\partial_r,\partial_s,\partial_t)
=i\partial_t+\partial_r^2+A\partial_r+h^{-2}\partial^2_s+B\partial_s+C,
\end{equation}
where $A=(2i\Phi_r+h^{-1}\kappa)$, $B=(2ih^{-2}\Phi_s+h^{-1}\partial_sh^{-1})$ 
and $C=i\triangle\Phi-(\nabla\Phi)^2$. The pseudo-differential approach, which uses 
a Nirenberg-type factorization~\cite{Nirenberg1973}, allows us to construct various order ABCs 
given by~\cite{AB2001,ABM2004}
\begin{align}
\ABC_{1a}:&\quad\partial_{{n}}u+e^{-i\pi/4}e^{i\Phi}(\partial_t-i\partial_s^2)^{1/2}e^{-i\Phi}u=0;\\
\ABC_{2a}:&\quad\partial_{{n}}u+e^{-i\pi/4}e^{i\Phi}(\partial_t-i\partial_s^2)^{1/2}e^{-i\Phi}u\nonumber\\
&\quad+\frac{1}{2}\kappa u+e^{-i\pi/4}\Phi_se^{i\Phi}
(\partial_t-i\partial_s^2)^{-1/2}\partial_s(e^{-i\Phi}u)
+\frac{1}{2}i\kappa
e^{i\Phi}(\partial_t-i\partial_s^2)^{-1}\partial_s^2(e^{-i\Phi}u)=0,
\end{align}
where 
\begin{equation}
(\partial_t-i\partial^2_s)^{1/2}f(s,t)=(\partial_t-i\partial^2_s)(\partial_t-i\partial^2_s)^{-1/2}f(s,t),
\end{equation}
assuming $f(s,0)=0$.

% \begin{itemize}[leftmargin=*]
% \item $\ABC_{1a}$:
% \begin{equation}
% \partial_{{n}}u+e^{-i\pi/4}e^{i\Phi}(\partial_t-i\partial_s^2)^{1/2}e^{-i\Phi}u=0;
% \end{equation}
% \item $\ABC_{2a}$:
% \begin{equation}
% \partial_{{n}}u+e^{-i\pi/4}e^{i\Phi}(\partial_t-i\partial_s^2)^{1/2}e^{-i\Phi}u\\
% +\frac{1}{2}\kappa u
% +e^{-i\pi/4}\Phi_se^{i\Phi}(\partial_t-i\partial_s^2)^{-1/2}\partial_s(e^{-i\Phi}u)\\
% +\frac{1}{2}i\kappa
% e^{i\Phi}(\partial_t-i\partial_s^2)^{-1}\partial_s^2(e^{-i\Phi}u)=0,
% \end{equation}
% where 
% \begin{equation}
% (\partial_t-i\partial^2_s)^{1/2}f(s,t)=(\partial_t-i\partial^2_s)(\partial_t-i\partial^2_s)^{-1/2}f(s,t),
% \end{equation}
% assuming $f(s,0)=0$.
% \end{itemize}
\begin{rmk}
For the field $u(\vv{x},t)$, the action of operators of the form
$(\partial_t-i\partial^2_s)^{\alpha},\alpha=1/2, -1/2, -1,\ldots,$ can be easily
computed by observing that the field can be given a periodic extension in
terms of the parametrization variable, $s$, so that the method discussed in
Remark~\ref{rmk:periodic} can be employed. The smoothness of the boundary is,
therefore, a necessary condition for this method to be applicable. From the 
previous sections, we know that the operator represented by
$(\partial_t-i\partial^2_s)^{-m/2}$, where $m$ is a positive integer, is defined as
\begin{equation}
(\partial_t-i\partial^2_s)^{-m/2}f(s,t)\\
=\frac{1}{2\pi}\iint{d\zeta}\,ds'\,e^{i\zeta (s-s')}
[\partial_{t'}^{-m/2}e^{-i\zeta^2(t-t')}{f}(s',t')]_{t'=t}.
\end{equation}
Let $\fourier_{s}f(s,t)=\tilde{f}(\zeta,t)$, then the operation $(\partial_t-i\partial^2_s)^{-m/2}$ 
involves the inverse Fourier transform of $\tilde{f}(\zeta,t')e^{-i\zeta^2(t-t')}$ 
with respect to $\zeta$. Therefore, if $f(s,t)$ is of the Schwartz class (with respect to $s\in\field{R}$) 
then so is $\tilde{f}(\zeta,t')e^{-i\zeta^2(t-t')}$, so that $(\partial_t-i\partial^2_s)^{-m/2}f(s,t)$ 
will also be of the Schwartz class (with respect to $s\in\field{R}$). Further, it is 
straightforward to show that $(\partial_t-i\partial^2_s)^{-m/2}f(s,t)$ is continuous at $t=0$. 

Next, our aim is to define the operator $(\partial_t-i\partial^2_s)^{-m/2}$ for 
tempered distributions $f(s,t)$ such that $\supp_tf\subset[0,\infty)$. To this end, 
let us observe that the transpose $[(\partial_t-i\partial^2_s)^{-m/2}]^{\tp}$ is given by
\begin{equation}
[(\partial_t-i\partial^2_s)^{-m/2}]^{\tp}g(s,t)\\
=\frac{1}{2\pi}\iint{d\zeta}\,ds'\,e^{i\zeta (s-s')}
[(\partial_{t'}^{-m/2})^{\tp}e^{-i\zeta^2(t-t')}{g}(s',t')]_{t'=t},
\end{equation}
where $(\partial_{t'}^{-m/2})^{\tp}$ denotes the Weyl fractional integral. The domain of 
definition of this operator is evident. Now, for any $g(s,t)\in\Schwartz(\field{R}\times\field{R})$ 
with $\supp_t\,g\subset[0,\infty)$, the operation $(\partial_t-i\partial^2_s)^{-m/2}f(s,t)$ 
for distributions $f(s,t)\in\Schwartz'(\field{R}\times\field{R})$ 
with $\supp_t\,f\subset[0,\infty)$ can be defined by introducing a Schwartz class function $g_1(s,t)$ 
such that it agrees with $[(\partial_t-i\partial^2_s)^{-m/2}]^{\tp}g(s,t)$ for $t\in[0,\infty)$ so that
\begin{equation}
\langle g,(\partial_t-i\partial^2_s)^{-m/2}f\rangle
=\langle [(\partial_t-i\partial^2_s)^{-m/2}]^{\tp}g,f\rangle
=\langle g_1,f\rangle.
\end{equation}
Therefore, $(\partial_t-i\partial^2_s)^{-m/2}f(s,t)$ is also tempered such that
$\supp_t\,(\partial_t-i\partial^2_s)^{-m/2}f(s,t)\subset[0,\infty)$. Note that
any function of the tangential variable, $s$, can be extended periodically on
the whole line. Thus, periodic functions being a tempered distribution 
can be easily included in the domain of definition of
$(\partial_t-i\partial^2_s)^{-m/2}$. It is interesting to observe that by applying 
the Leibniz formula for the fractional integrals one can obtain local 
approximation of the ABCs with respect to $s$:
\begin{widetext}
\begin{equation}
\begin{split}
(\partial_t-i\partial^2_s)^{-m/2}f(s,t)
&=\frac{1}{2\pi}\iint{d\zeta}\,ds'\,e^{i\zeta (s-s')}
[\partial_{t'}^{-m/2}e^{-i\zeta^2(t-t')}{f}(s',t')]_{t'=t}\\
&=\frac{1}{2\pi}\iint{d\zeta}\,ds'\,e^{i\zeta (s-s')}
\left[\sum_{j\in\field{N}}\binom{-m/2}{j}(\partial^j_{t'}e^{-i\zeta^2(t-t')})
\partial_{t'}^{-m/2-j}{f}(s',t')\right]_{t'=t}\\
&=\frac{1}{2\pi}\iint{d\zeta}\,ds'\,e^{i\zeta (s-s')}
\left[\sum_{j\in\field{N}}\binom{-m/2}{j}(-i\zeta^{2})^j\partial_{t}^{-m/2-j}{f}(s',t)\right]\\
&=\sum_{j\in\field{N}}\binom{-m/2}{j}(i\partial_s^{2})^j\partial_{t}^{-m/2-j}{f}(s,t).
\end{split}
\end{equation}
\end{widetext}
\end{rmk}
The second family of various order ABCs are obtained under high-frequency assumption with respect to the
temporal frequency. These are given by
\begin{align}
\ABC_{1b}:&\quad\partial_{{n}}u+e^{-i\pi/4}e^{i\Phi}\partial_t^{1/2}e^{-i\Phi}u=0;\\
\ABC_{2b}:&\quad\partial_{{n}}u+e^{-i\pi/4}e^{i\Phi}\partial_t^{1/2}e^{-i\Phi}u 
+\frac{1}{2}\kappa u-\frac{1}{2}e^{i\pi/4}\left(\frac{\kappa^2}{4}
+\partial_s^2\right)e^{i\Phi}\partial_t^{-1/2}e^{-i\Phi}u=0;\\
\ABC_{3b}:&\quad\partial_{{n}}u+e^{-i\pi/4}e^{i\Phi}\partial_t^{1/2}e^{-i\Phi}u
+\frac{1}{2}\kappa u-\frac{1}{2}e^{i\pi/4}\left(\frac{\kappa^2}{4}
+\partial_s^2\right)e^{i\Phi}\partial_t^{-1/2}e^{-i\Phi}u\nonumber\\
&\qquad+\frac{i}{2}\left[\frac{\kappa^3+\partial_s^2\kappa}{4}
-\frac{\partial_{{n}}\phi}{2}+\partial_s(\kappa\partial_s)\right]e^{i\Phi}\partial^{-1}_te^{-i\Phi}u=0.
\end{align}

% \begin{itemize}[leftmargin=*]
% \item $\ABC_{1b}$:
% \begin{equation}
% \partial_{{n}}u+e^{-i\pi/4}e^{i\Phi}\partial_t^{1/2}e^{-i\Phi}u=0;
% \end{equation}
% \item $\ABC_{2b}$:
% \begin{equation}
% \partial_{{n}}u+e^{-i\pi/4}e^{i\Phi}\partial_t^{1/2}e^{-i\Phi}u+\frac{1}{2}\kappa u\\
% -\frac{1}{2}e^{i\pi/4}\left(\frac{\kappa^2}{4}
% +\partial_s^2\right)e^{i\Phi}\partial_t^{-1/2}e^{-i\Phi}u=0;
% \end{equation}
% \item $\ABC_{3b}$:
% \begin{equation}
% \partial_{{n}}u+e^{-i\pi/4}e^{i\Phi}\partial_t^{1/2}e^{-i\Phi}u+\frac{1}{2}\kappa u\\
% -\frac{1}{2}e^{i\pi/4}\left(\frac{\kappa^2}{4}+\partial_s^2\right)e^{i\Phi}\partial_t^{-1/2}e^{-i\Phi}u\\
% +\frac{i}{2}\left[\frac{\kappa^3+\partial_s^2\kappa}{4}
% -\frac{\partial_{{n}}\phi}{2}+\partial_s(\kappa\partial_s)\right]e^{i\Phi}\partial^{-1}_te^{-i\Phi}u=0.
% \end{equation}
% \end{itemize}
%======================================================================================
\subsubsection{Stability and uniqueness}
In order to study the stability property of the solution of the IVP defined 
by~\eqref{eq:2D-NL} with boundary condition $\ABC_{1a}$ and $\ABC_{2a}$,
respectively, we start with the relation:
\begin{equation*}
\|u(\cdot,T)\|^2_{\fs{L}^2(\Omega_i)}-\|u_0\|^2_{\fs{L}^2(\Omega_i)}
=2\Re\int_0^T dt\int_{\Gamma}(u^*i\nabla u)\cdot d\vbs{\varsigma}.
\end{equation*}
Realization of the boundary operators require an appropriate Fourier
representation with respect to the tangential variable, $s$. To this end, one
may either employ the Fourier series representation by extending the field 
periodically for all $s\in\field{R}$ or Fourier transform representation by 
extending the field to all $s\in\field{R}$ by setting it zero outside $\Gamma$. 
The result for the first order ABCs, $\ABC_{1a}$, can be obtained by observing
that the boundary integral
\begin{equation}
\mathcal{I}_1 = -e^{i\pi/4}\int_0^Tdt\int_{\Gamma}
\overline{(ue^{-i\Phi})}\sqrt{\partial_t-i\partial^2_s}(u e^{-i\Phi})ds,
\end{equation}
satisfies $\Re\mathcal{I}_1\leq0$ so that
\begin{equation*}
\|u(\cdot,T)\|_{\fs{L}^2(\Omega_i)}\leq\|u_0\|_{\fs{L}^2(\Omega_i)}.
\end{equation*}
For the second order ABCs, $\ABC_{2a}$, the energy estimate cannot be
obtained for a general potential function. A special case of interest is when 
$\partial_s\phi=0$ so that $\ABC_{2a}$ is given by
\begin{equation}
\partial_{{n}}u+e^{-i\pi/4}e^{i\Phi}(\partial_t-i\partial_s^2)^{1/2}e^{-i\Phi}u+\frac{1}{2}\kappa u\\
+\frac{1}{2}i\kappa e^{i\Phi}(\partial_t-i\partial_s^2)^{-1}\partial_s^2(e^{-i\Phi}u)=0.
\end{equation}
Define the boundary integrals
\begin{equation}
\begin{split}
&\mathcal{I}_2 = -\frac{i}{2}\int_0^Tdt\int_{\Gamma}\kappa(s)|u|^2ds,\\
&\mathcal{I}_3 = \frac{1}{2}\int_0^Tdt\int_{\Gamma}
\kappa(s)\overline{(ue^{-i\Phi})}(\partial_t-i\partial^2_s)^{-1}
\partial_s^2(ue^{-i\Phi})ds.
\end{split}
\end{equation}
It follows that $\Re\mathcal{I}_2=0$. Define the symbol 
$\sigma_{P}=(1/2)\kappa(s)(i\xi+i\zeta^2)^{-1}(-\zeta^2)$ so that the
symbol for the adjoint operator $\sigma_{P^{\dagger}}$ works out to be
\begin{equation*}
\sigma_{P^{\dagger}}\sim\frac{-i}{2}\sum_{k\in\field{N}}\frac{1}{k!i^k}(\partial_s^k\kappa)
\partial^k_{\zeta}\left(\frac{\zeta^2}{\xi+\zeta^2}\right).
\end{equation*}
Let $2Q=P+P^{\dagger}$ so that 
\begin{equation*}
\begin{split}
2\sigma_{Q}
&\sim\frac{i}{2}\frac{\kappa\zeta^2}{\xi+\zeta^2}+\frac{-i}{2}
\sum_{k\in\field{N}}\frac{1}{k!i^k}(\partial_s^k\kappa)\partial^k_{\zeta}
\left(\frac{\zeta^2}{\xi+\zeta^2}\right)\\
&=-\frac{1}{2}(\partial_s\kappa)\partial_{\zeta}
\left(\frac{\zeta^2}{\xi+\zeta^2}\right)+\ldots,
\end{split}
\end{equation*}
and putting $\psi=u e^{-i\Phi}$, we have
\begin{equation*}
\Re\mathcal{I}_3=\frac{1}{(2\pi)^2}\iiint d\zeta d\xi ds
\fourier_{t}\fourier_{s}[\psi(s,t)]
\sigma_{Q}(s,\zeta,\xi)\fourier_t[\psi(s,t)]e^{-i\zeta s}.
\end{equation*}
From the asymptotic expansion, to the zeroth order, we have
$\Re\mathcal{I}_3\approx 0$ (for constant curvature boundary, this result is
exact). The energy estimate can thus be approximately established for the
special case when $\partial_s\phi=0$.

For the ABCs obtained in the high-frequency approximation, it is evident that
the first order ABC, $\ABC_{1b}$, satisfies the energy estimate. For the
higher-order ABCs, one requires an additional conditions, $\partial_s\phi=0$ in
order to obtain the energy estimate. In $\ABC_{2b}$, the term with the factor
$\kappa/2$ can be ignored as it leads to a purely imaginary quantity. 
Putting $\psi=ue^{i\Phi}$, consider the following boundary integrals
\begin{equation}
\begin{split}
&\mathcal{I}_1 =
-\frac{e^{i\pi/4}}{2}\int_0^Tdt\int_{\Gamma}\psi^*(s,t)
\left[2\partial_t^{\frac{1}{2}}-i\partial^{-\frac{1}{2}}_t\partial_s^2\right]\psi(s,t)ds,\\
&\mathcal{I}_2 =
-\frac{e^{-i\pi/4}}{8}\int_0^Tdt
\int_{\Gamma}\psi^*(s,t)\kappa^2\partial^{-\frac{1}{2}}_t\psi(s,t)ds.
\end{split}
\end{equation}
The first integral, $\mathcal{I}_1$, can be dealt with in the manner done before. The second
integral can be dealt with in the manner described in the appendix. Defining 
$\sigma_P=-e^{-i\pi/4}\kappa^2(i\xi)^{-1/2}$ so that
$\sigma_{P^{\dagger}}=-e^{i\pi/4}\kappa^2(-i\xi)^{-1/2}$. Putting
$2Q=P+P^{\dagger}$, we have $\sigma_Q=-\kappa^2\cos[\pi/4+\pi\sgn(\xi)/4]$.
The boundary integral can now be written as 
\begin{equation*}
\Re\mathcal{I}_2=-\frac{1}{16\pi}\iint d\xi ds
[\kappa(s)]^2|\xi|^{-1/2}\cos[\pi(1+\sgn(\xi))/4]|\fourier_t[\psi(s,t)]|^2,
\end{equation*}
so that $\Re\mathcal{I}_2\leq0$.

For $\ABC_{3b}$, we define the following boundary integrals
\begin{equation}
\begin{split}
&\mathcal{I}_3 =
\frac{1}{8}\int_0^Tdt\int_{\Gamma}\psi^*(s,t)
[(\kappa^3+\partial_s^2\kappa)\partial_t^{-1}]\psi(s,t)ds,\\
&\mathcal{I}_4 =
\frac{1}{2}\int_0^Tdt\int_{\Gamma}\psi^*(s,t)
[(\partial_s\kappa\partial_s)\partial_t^{-1}]\psi(s,t)ds,\\
&\mathcal{I}_5 =
-\frac{1}{4}\int_0^Tdt\int_{\Gamma}\psi^*(s,t)
[(\partial_n\phi)\partial_t^{-1}]\psi(s,t)ds.
\end{split}
\end{equation}
Defining $\sigma_P=(\kappa^3+\partial_s^2\kappa)(i\xi)^{-1}$ so that
\[
\sigma_Q=(\kappa^3+\partial_s^2\kappa)|\xi|^{-1}\cos[\pi\sgn(\xi)/2].
\] 
From the above equation, it follows that $\mathcal{I}_3$ is purely imaginary. Carrying out an
integration by parts in $\mathcal{I}_4$ with respect to $s$, we have
\begin{equation*}
\begin{split}
\mathcal{I}_4
&=-\frac{1}{2}\int_0^Tdt\int_{\Gamma}[\partial_s\psi^*(s,t)]\kappa(s)\partial_t^{-1}[\partial_s\psi(s,t)]ds\\
&=-\frac{1}{2\pi}\int_0^Td\xi\int_{\Gamma}\kappa(s)(i\xi)^{-1}|\fourier_t[\partial_s\psi(s,t)]|^2ds.
\end{split}
\end{equation*}
Therefore, $\mathcal{I}_4$ is purely imaginary. The discussion of the last
integral, $\mathcal{I}_5$, is more involved. Define the symbol 
$\sigma_{P}=-(\partial_n\phi)(i\xi)^{-1}$ so that the
symbol for the adjoint operator $\sigma_{P^{\dagger}}$ works out to be
\begin{equation*}
\sigma_{P^{\dagger}}\sim-\sum_{k\in\field{N}}\frac{1}{k!i^k}(\partial_t^k\partial_n\phi)
\partial^k_{\xi}(-i\xi)^{-1}.
\end{equation*}
Now
\begin{equation*}
\begin{split}
2\sigma_{Q}
&\sim-(\partial_n\phi)(i\xi)^{-1}-\sum_{k\in\field{N}}\frac{1}{k!i^k}
(\partial_t^k\partial_n\phi)\partial^k_{\xi}(-i\xi)^{-1}\\
&=-(\partial_t\partial_n\phi)(\xi)^{-2}+\ldots.
\end{split}
\end{equation*}
Therefore, the contribution from this last integral can be ignored on account
of the lower order leading term in the above asymptotic expansion. Note that an
alternative technique is to symmetrize the operator corresponding to
$\mathcal{I}_5$ as suggested in Ref.~\onlinecite{ABK2012}:
\begin{equation*}
-\frac{i\sgn[\partial_{n}\phi]}{4}\sqrt{|\partial_{n}\phi|}e^{i\Phi}
\partial^{-1}_t\left(\sqrt{|\partial_{n}\phi|}e^{-i\Phi}u\right)\\
=-\frac{i\partial_{{n}}\phi}{4}e^{i\Phi}\partial^{-1}_te^{-i\Phi}u
\mod{(\partial_t^{-2})}.
\end{equation*}
While the order of the error term is the same for both alternatives, the unsymmetrized and
the symmetrized version, the numerical conditioning of the symmetrized version is
superior because $\mathcal{I}_5$ becomes purely imaginary~\cite{ABK2012}. Thus the energy estimate
\begin{equation*}
\|u(\cdot,T)\|_{\fs{L}^2(\Omega_i)}\leq\|u_0\|_{\fs{L}^2(\Omega_i)}.
\end{equation*}
can be established (at least in the ``weak'' sense) for the ABC-family:
$\ABC_{jb},j=1,2,3$. Let us note that only the second and third order ABCs require 
$\partial_s\phi=0$ in order to establish the energy estimate.

\subsection{Domains with straight boundary: infinite strip }
The ABCs for the straight boundary can be written by setting the curvature $\kappa$ 
to zero:
\begin{align}
\ABC_{1a}:&\quad\partial_{{n}}u+e^{-i\pi/4}e^{i\Phi}(\partial_t
-i\partial_s^2)^{1/2}e^{-i\Phi}u=0;\\
\ABC_{2a}:&\quad\partial_{{n}}u+e^{-i\pi/4}e^{i\Phi}(\partial_t-i\partial_s^2)^{1/2}e^{-i\Phi}u
+e^{-i\pi/4}\Phi_se^{i\Phi}(\partial_t-i\partial_s^2)^{-1/2}\partial_s(e^{-i\Phi}u)=0.
\end{align}
Along the direction in which the strip extends to infinity, we impose the periodic 
boundary condition. Note that an exact representation of the boundary operators is 
possible in this case which is based on the observations made in earlier sections. 
The energy estimate for the first order ABCs is easy to obtain; however, for the 
second order ABCs this result is not available. The high-frequency ABCs work out to be
\begin{align}
\ABC_{1b}:&\quad\partial_{{n}}u+e^{-i\pi/4}e^{i\Phi}\partial_t^{1/2}e^{-i\Phi}u=0;\\
\ABC_{2b}:&\quad\partial_{{n}}u+e^{-i\pi/4}e^{i\Phi}\partial_t^{1/2}e^{-i\Phi}u
-\frac{1}{2}e^{i\pi/4}\partial_s^2\left(e^{i\Phi}\partial_t^{-1/2}e^{-i\Phi}u\right)=0;\\
\ABC_{3b}:&\quad\partial_{{n}}u+e^{-i\pi/4}e^{i\Phi}\partial_t^{1/2}e^{-i\Phi}u
-\frac{1}{2}e^{i\pi/4}\partial_s^2\left(e^{i\Phi}\partial_t^{-1/2}e^{-i\Phi}u\right)
-\frac{i\partial_{{n}}\phi}{4}e^{i\Phi}\partial^{-1}_te^{-i\Phi}u=0.
\end{align}
The energy estimate for the general case is not available for such ABCs.

\subsection{Rectangular domains: corner conditions}
Continuing with the rectangular domain as defined in Sec.~\ref{sec:rect-domain}, let 
us consider the possibility of extending the results
obtained in Sec.~\ref{sec:HF-rect} to the nonlinear/variable potential case. 
We confine our attention to the high-frequency ABCs only. We demonstrate the
possibility of constructing the corner condition at the corner defined by 
$\Gamma_r\cap\Gamma_t$. Consider the $\ABC_{2b}$ given by
\begin{equation}
\partial_{x_1}u+e^{-i\pi/4}e^{i\Phi}\partial_t^{1/2}e^{-i\Phi}u\\
-\frac{1}{2}e^{i\pi/4}\partial_{x_2}^2
\left(e^{i\Phi}\partial_t^{-1/2}e^{-i\Phi}u\right)=0,\quad\vv{x}\in\Gamma_r,
\end{equation}
and
\begin{equation}
\partial_{x_2}u+e^{-i\pi/4}e^{i\Phi}\partial_t^{1/2}e^{-i\Phi}u\\
-\frac{1}{2}e^{i\pi/4}\partial_{x_1}^2\left(e^{i\Phi}\partial_t^{-1/2}e^{-i\Phi}u\right)
=0,\quad\vv{x}\in\Gamma_t.
\end{equation}
Let us consider the weak formulation of the IVP given by
\begin{equation}\label{eq:IVP-NL-estimate}
\begin{split}
&i\partial_tu+\triangle u+\phi u=0,\quad(\vv{x},t)\in\Omega_i\times\field{R}_+,\\
&u(\vv{x},0)=u_0(\vv{x}),\quad\vv{x}\in\Omega_i,\quad\supp\,u_0\subset\Omega_i.
\end{split}\end{equation}
Let $\psi(\vv{x})\in \fs{W}^{1,2}(\Omega_i)$ be a test function so that
\begin{equation}
\int_{\Omega_i}(i\partial_tu+\nabla^2u+\phi u)\psi d^2\vv{x}\\
=\int_{\Omega_i}[i\partial_tu-(\nabla u)\cdot(\nabla\psi)+\phi u]d^2\vv{x}
+\int_{\Gamma}\psi(\nabla u)\cdot d\vbs{\varsigma}.
\end{equation}
Focusing on the top and right boundary, the boundary integrals are given by
\begin{widetext}
\begin{equation}
\begin{split}
%\mathcal{I}_R+\mathcal{I}_T
&\int_{\Gamma_r}\psi(\partial_{{x_1}}u) dx_2
+\int_{\Gamma_t}\psi(\partial_{{x_2}}u) dx_1\\
&=-e^{-i\pi/4}\int_{\Gamma_r\cup\Gamma_l}\psi e^{i\Phi}\partial^{1/2}_te^{-i\Phi}u
+\frac{1}{2}e^{i\pi/4}\left[\int_{\Gamma_r}\psi\partial^2_{{x_2}}
\left(e^{i\Phi}\partial^{-1/2}_te^{-i\Phi}u\right) dx_2
+\int_{\Gamma_t}\psi\partial^2_{{x_1}}
\left(e^{i\Phi}\partial^{-1/2}_te^{-i\Phi}u\right)dx_1\right]\\
&=-e^{-i\pi/4}\int_{\Gamma_r\cup\Gamma_l}\psi e^{i\Phi}\partial^{1/2}_te^{-i\Phi}u
+\frac{1}{2}e^{i\pi/4}\left[
\left.\psi\partial_{x_2}\left(e^{i\Phi}\partial^{-1/2}_te^{-i\Phi}u\right)\right|_{x_2=x_b}^{x_t}
+\left.\psi\partial_{x_1}\left(e^{i\Phi}\partial^{-1/2}_te^{-i\Phi}u\right)\right|_{x_1=x_l}^{x_r}\right]\\
&\quad-\frac{1}{2}e^{i\pi/4}\left[
\int_{\Gamma_r}(\partial_{x_2}\psi)\partial_{{x_2}}
\left(e^{i\Phi}\partial^{-1/2}_te^{-i\Phi}u\right) dx_2
+\int_{\Gamma_t}(\partial_{x_1}\psi)\partial_{{x_1}}
\left(e^{-i\Phi}\partial^{-1/2}_te^{-i\Phi}u\right)dx_1\right].
\end{split}
\end{equation}
\end{widetext}
From here, it is evident that the corner condition must provide the value of the following expression at 
$\Gamma_r\cap\Gamma_t$:
\begin{multline}
\partial_{x_1}\left(e^{i\Phi}\partial_t^{-1/2}e^{-i\Phi}u\right)
+\partial_{x_2}\left(e^{i\Phi}\partial_t^{-1/2}e^{-i\Phi}u\right)\\
=i(\Phi_{x_1}+\Phi_{x_2})e^{i\Phi}\partial_t^{-1/2}e^{-i\Phi}u
-ie^{i\Phi}\partial_t^{-1/2}\left[(\Phi_{x_1}+\Phi_{x_2})e^{-i\Phi}u\right]
+e^{i\Phi}\partial_t^{-1/2}e^{-i\Phi}(\partial_{x_1}u+\partial_{x_2}u).
\end{multline}
For the linear case with variable potential, the last term in the above 
equation must be computed from a corner condition. However, for the nonlinear 
case, $\Phi$ is supposed to be dependent on some known field while our 
final intent is to restore its nonlinear dependence. In this light, the 
above equation cannot be made free of the derivatives of $u$ at the corner
as the equation is nonlinear. Nevertheless, we may derive a condition at the corner
which can be combined with the ABCs and eventually be solved by an iterative 
scheme. Putting $u=\psi e^{i\Phi}$ in the evolution 
equation~\eqref{eq:IVP-NL-estimate}, we have 
\begin{equation}
i\partial_t\psi+\triangle\psi+2i(\nabla\Phi)\cdot(\nabla\psi)+(e^{-i\Phi}\triangle
e^{i\Phi})\psi=0.
\end{equation}
Carrying out the operation $e^{i\Phi}\partial_t^{-1/2}$ on both side of the
equation above, we have
\begin{widetext}
\begin{equation}
\begin{split}
&ie^{i\Phi}\partial_t^{1/2}\psi+e^{i\Phi}\partial_t^{-1/2}\triangle\psi
+2e^{i\Phi}\partial_t^{-1/2}(i\nabla\Phi)\cdot(\nabla\psi)
+e^{i\Phi}\partial_t^{-1/2}(e^{-i\Phi}\triangle e^{i\Phi})\psi=0,\\
&ie^{i\Phi}\partial_t^{1/2}\psi+e^{i\Phi}\partial_t^{-1/2}\triangle\psi
+2e^{i\Phi}(i\nabla\Phi)\cdot\partial_t^{-1/2}(\nabla\psi)+(\triangle 
e^{i\Phi})\partial_t^{-1/2}\psi+\ldots=0,\\
\end{split}
\end{equation}
\end{widetext}
which simplifies to
\begin{equation}
ie^{i\Phi}\partial_t^{1/2}e^{-i\Phi}u
+\triangle\left(e^{i\Phi}\partial_t^{-1/2}e^{-i\Phi}u\right)
=0\,\,\mod(\partial_t^{-3/2}e^{-i\Phi}u).
\end{equation}
This equation combined with the ABCs gives the following corner condition
\begin{equation}
\text{CC}_1:\,\,\partial_{x_1}u
+\partial_{x_2}u+\frac{3}{2}e^{-i\pi/4}e^{i\Phi}\partial_t^{1/2}e^{-i\Phi}u=0,
\end{equation}
or, equivalently,
\begin{equation}
\text{CC}_1:\,\,e^{i\Phi}\partial_t^{-1/2}e^{-i\Phi}(\partial_{x_1}u
+\partial_{x_2}u)+\frac{3}{2}e^{-i\pi/4}u=0.
\end{equation}
Similarly, for $\ABC_{3b}$ one has the following corner condition
\begin{equation}
\text{CC}_2:\,\,\partial_{x_1}u+\partial_{x_2}u
+\frac{3}{2}e^{-i\pi/4}e^{i\Phi}\partial_t^{1/2}e^{-i\Phi}u\\
-i\frac{\partial_{x_1}\phi+\partial_{x_2}\phi}{4}e^{i\Phi}\partial^{-1}_te^{-i\Phi}u
=0\,\,\mod(\partial^{-5/2}_te^{-i\Phi}u),
\end{equation}
or, equivalently,
\begin{equation}
\text{CC}_2:\,\,e^{i\Phi}\partial_t^{-1/2}e^{-i\Phi}(\partial_{x_1}u
+\partial_{x_2}u)+\frac{3}{2}e^{-i\pi/4}u\\
-i\frac{\partial_{x_1}\phi+\partial_{x_2}\phi}{4}e^{i\Phi}\partial^{-3/2}_te^{-i\Phi}u
=0\,\,\mod(\partial^{-5/2}_te^{-i\Phi}u).
\end{equation}
\subsection{Special case: $\phi=\phi(t)$}
For time-dependent potentials with no spatial variation, the quantity $\Phi$ is
purely time-dependent and the ABCs for different domains types can be simplified
considerably. It must be remarked that the exact form of the transparent boundary condition
is only obtainable for an infinite-strip with periodic boundary condition in
the unbounded direction or rectangular domains. At the appropriate segments of
the boundary, it reads as
\begin{equation}
\partial_{{n}}u+e^{-i\pi/4}e^{i\Phi}(\partial_t-i\partial_s^2)^{1/2}e^{-i\Phi}u=0.
\end{equation}
For rectangular domains, we consider this problem in more detail. At the boundary
$\Gamma_r$, we have
\begin{equation}
\partial_{x_1}u+e^{-i\pi/4}e^{i\Phi}(\partial_t-i\partial_{x_2}^2)^{1/2}e^{-i\Phi}u=0.
\end{equation}
The auxiliary function, $\varphi(x_1,x_2,t,t')$ as defined in 
Sec.~\ref{sec:rect-domain} in the present case is given by
\begin{equation}
\fourier_{x_2}[\varphi(x_1,x_2,t,t')] =
e^{-i\zeta^2_2(t-t')-i\Phi(t')}\fourier_{x_2}[u(x_1,x_2,t')],
\end{equation}
so that $\varphi(x_1,x_2,t,t)=u(x_1,x_2,t)e^{-i\Phi(t)}$. Putting 
$\tilde{u}_0(\vbs{\zeta})=\fourier_{(x_1,x_2)}u_0(\vv{x})$, we have
\begin{equation}
\begin{split}
&\fourier_{(x_1,x_2)}u(\vv{x},t')=e^{-i(\zeta_1^2+\zeta_2^2)t'+i\Phi(t')}
\tilde{u}_0(\vbs{\zeta}),\\
&\varphi(x_1,x_2,t,t')=\frac{1}{(2\pi)^2}\int_{\field{R}^2}
e^{i\vbs{\zeta}\cdot\vv{x}-i\zeta_1^2t'-i\zeta_2^2t}
\tilde{u}_0(\vbs{\zeta})d^2\vbs{\zeta}.
\end{split}
\end{equation}
From here is follows that the IVPs satisfied by this auxiliary field is not
different from that described in Sec.~\ref{sec:rect-domain}.

\section{Conclusion}
In this paper, we have discussed the formulation of the operator
$(\partial_t-i\triangle_{\Gamma})^{\alpha}, \alpha=1/2,-1/2,-1,\ldots,$ in terms of the
fractional operators in various settings. This allows the TBCs/ABCs for the free Schr\"odinger equation 
and general Schr\"odinger equation formulated on various types of computational
domains to be expressed in a natural way. In particular, two families of ABCs
within the gauge transformation strategy are studied in this paper: the ABCs obtained with the 
pseudo-differential approach and those obtained as a 
high-frequency approximation of the former. For the rectangular
domains, we have developed various order corner conditions for the family of
ABCs obtained in the high-frequency approximation. Each of these families of 
ABCs (along with corner conditions) are also investigated for stability and
uniqueness of the solution of the resulting initial-boundary value problem.
Further, we expect that the results presented in this paper can be easily extended to 
the pseudo-differential approach within the direct strategy~\cite{ABK2012}. Finally, 
let us remark that the ABCs obtained in this article can be readily discretized 
using the convolution quadrature. These issues will be addressed in a forthcoming paper.

\appendix
\section{Some properties of the pseudo-differential operators\label{app:psido}}
Let us consider the symbol space $\fs{S}^m_M(Y\times\field{R}^n)$ of $M$-quasi
homogeneous symbols~\cite{Lascar1977} where $Y$ is an open subset of $\field{R}^n$ and
$M=(\mu_1,\mu_2,\ldots,\mu_n)$ is an n-tuple of numbers $\mu_i>0$. Let $P$ be a 
pseudo-differential operator with the symbol
$p(y,\zeta)\in\fs{S}_M^m(Y\times\field{R}^n)$ so that
\begin{equation}
\begin{split}
Pu(y)&=\int p(y,\zeta)e^{i\zeta\cdot y}\tilde{u}(\zeta)d^n\zeta\\
&=\iint p(y,\zeta)e^{i\zeta\cdot(y-y')}u(y')dy'd\zeta,
\end{split}
\end{equation}
where $u(y)\in{\fs{C}_0^{\infty}(Y)}$ and
$d\zeta=d\zeta_1 d\zeta_2\ldots,d\zeta_n$, $dy=dy_1 dy_2\ldots,dy_n$. The adjoint of 
this operator, denoted by ${P}^{\dagger}$, can be defined as
\begin{equation}
P^{\dagger} u(y)=(2\pi)^{-n}\iint p^*(y',\zeta)e^{i\zeta\cdot(y-y')}u(y')dy'd\zeta.
\end{equation}
This operator belongs to a more general class of pseudo-differential operators defined by
\begin{equation}
Pu(y)=\iint p(y,y',\zeta)e^{i\zeta\cdot(y-y')}u(y')dy'd\zeta,
\end{equation}
where the symbol $p(y,y',\zeta)\in\fs{C}^{\infty}(Y\times Y\times\field{R}^n)$ 
lies in $\fs{S}_M^m(Y\times Y\times\field{R}^n)$. The adjoint of this operator is defined by the symbol
\begin{equation}
p^{\dagger}(y,y',\zeta)=p^*(y',y,\zeta).
\end{equation}
Note that the primed variable is to be integrated over. These new operators do not have unique symbols but 
they do admit of a representation in terms of the former kind of operators as 
$P=\OP(\sigma_P(y,\zeta))+R$ where $\sigma_P(y,\zeta)\in\fs{S}_M^m(Y\times\field{R}^n)$ and 
$R$ is an operator with kernel $K_R(y,y')\in\fs{C}^{\infty}(Y\times Y)$ such that
%\begin{widetext}
\begin{equation}\label{eq:asymptotic-defn}
\sigma_P(y,\zeta)-\sum_{\alpha\in\field{N}^n,\,{|\alpha|}<N-1}
\frac{1}{{\alpha!}i^{|\alpha|}}\partial^{\alpha}_{\zeta}
[\partial^{\alpha}_{y'}p(y,y',\zeta)]_{y'=y}\\
\in\fs{S}^{m-\mu N}_M(Y\times Y\times\field{R}^n),
\end{equation}
%\end{widetext}
where $N>0$ and $\mu$ is smallest element of $M$. The relationship in 
equation~\eqref{eq:asymptotic-defn} defines the asymptotic expansion of the symbol and we write
\begin{equation}\label{eq:asymptotic-expansion}
\sigma_P(y,\zeta)\sim\sum_{\alpha\in\field{N}^n}
\frac{1}{\alpha!i^{|\alpha|}}\partial^{\alpha}_{\zeta}[\partial^{\alpha}_{y'}p(y,y',\zeta)]_{y'=y}.
\end{equation}
This determines $\sigma_P(y,\zeta)$ only up to a smoothing operator. The formula
expressing $\sigma_{P^{\dagger}}(y,\zeta)$ in terms of $\sigma^*_P(y',\zeta)$ can be obtained by writing
\begin{equation}
p^*(y',y,\zeta)-\sigma_P^*(y',\zeta) \in\fs{S}^{-\infty}_M(Y\times Y\times\field{R}^n),,
\end{equation}
and using the asymptotic expansion~\eqref{eq:asymptotic-expansion}
\begin{equation}
\begin{split}
\sigma_{P^{\dagger}}(y,\zeta)&\sim\sum_{\alpha\in\field{N}^n}\frac{1}{\alpha!i^{|\alpha|}}\partial^{\alpha}_{\zeta}[\partial^{\alpha}_{y'}\sigma^*_P(y',\zeta)]_{y'=y},
\end{split}
\end{equation}
where $R$ being a smoothing operator does not show up in the asymptotic
expansion. The adjoint can 
be used to write the Fourier transform of $Pu(y)$ by noticing that
$(P^{\dagger})^{\dagger}=P$ so that
\begin{equation}
Pu(y)=(2\pi)^{-n}\int\left(\int\sigma^*_{P^{\dagger}}(y',\zeta)
e^{-i\zeta\cdot y'}u(y')dy'\right)e^{i\zeta\cdot y}d\zeta.
\end{equation}
Consider the inner product defined by $\langle u|v\rangle = \int u^*v dy$. An 
expression of the form $\Re\langle u|P u\rangle$ arises in establishing the
stability of the ABCs. This can be computed by observing
\begin{equation}
2\Re\langle u|P u\rangle=\langle u|P u\rangle+\langle u|P^{\dagger} u\rangle.
\end{equation}
Define $2Q = (P+P^{\dagger})$ and using Plancheral's theorem, we have
\begin{equation}
\Re\langle u|P u\rangle=(2\pi)^{-n}\iint\,d\zeta dy'
\tilde{u}^*(\zeta)\sigma_{Q}(y',\zeta)u(y')e^{-i\zeta\cdot y'}.
\end{equation}
If $\sigma_{Q}(y,\zeta)$ is independent of $y$, then 
\begin{equation}
\Re\langle u|P u\rangle=(2\pi)^{-n}\int\,d\zeta|\tilde{u}(\zeta)|^2\sigma_{Q}(\zeta).
\end{equation}
The sign of this expression can solely be decided by the sign of
$\sigma_{Q}(\zeta)$. If $\sigma_{Q}(y,\zeta)=\phi(y_1)\sigma(\tilde{\zeta})$
where $\tilde{y}=(y_2,y_3,\ldots,y_n)\in\field{R}^{n-1}$ and
$\tilde{\zeta}=(\zeta_2,\zeta_3,\ldots,\zeta_n)\in\field{R}^{n-1}$ then
\begin{equation}
\Re\langle u|P u\rangle=(2\pi)^{-(n-1)}\iint\,
dy_1d\tilde{\zeta}|\fourier_{\tilde{y}}u(y_1,y)|^2\phi(y_1)\sigma(\tilde{\zeta}).
\end{equation}
On account of the ambiguity in the knowledge of the exact
symbol, the conclusion remains valid only up to an infinitely smoothing
operator (except for the special cases where the symbol is of principle type).
Some relevant examples are discussed below.
\begin{itemize}[leftmargin=*]
\item Fractional operators with symbol $p(t,\xi)=\eta(i\xi)^{\alpha}$ 
(where $t\in\field{R}$ with covariable $\xi\in\field{R}\setminus\{0\}$): 
\begin{equation}
\begin{split}
\sigma_{Q}(\xi)
&\sim\frac{1}{2}[\eta(i\xi)^{\alpha}+\eta^*(-i\xi)^{\alpha}]\\
&=|\eta||\xi|^{\alpha}\cos\left[\frac{\pi\alpha}{2}\sgn(\xi)+\arg{\eta}\right].
\end{split}
\end{equation}
\item Let $\sigma_P=\eta\phi(x)(i\xi)^{\alpha}$ with real valued function
$\phi(x)>0,\forall x\in\field{R}$ where the meaning
of the variables is same as that of the last example. Then
\begin{equation}
\begin{split}
\sigma_{Q}(x,t,\zeta,\xi)
&\sim\frac{1}{2}[\eta\phi(x)(i\xi)^{\alpha}+\eta^*\phi(x)(-i\xi)^{\alpha}]\\
&=|\eta|\phi(x)|\xi|^{\alpha}\cos\left[\frac{\pi\alpha}{2}\sgn(\xi)+\arg{\eta}\right],
\end{split}
\end{equation}
so that
\begin{equation}
\Re\langle u|P u\rangle=\frac{|\eta|}{2\pi}\iint\,d\xi dx|
\fourier_t[{u}(x,t)](x,\xi)|^2\phi(x)|\xi|^{\alpha}\times\\
\cos\left[\frac{\pi\alpha}{2}\sgn(\xi)+\arg{\eta}\right].
\end{equation}
\item Operators with symbol $p(x,t,\zeta,\xi)=\eta(i\xi+i \zeta^2)^{\alpha}$ 
(where $(x,t)\in\field{R}^2$ with covariables 
$(\zeta,\xi)\in\field{R}^2\setminus\{(\xi,\zeta)\in\field{R}^2:\xi+\zeta^2=0\}$): 
\begin{equation}
\begin{split}
\sigma_{Q}(\zeta,\xi)
&\sim\frac{1}{2}[\eta(i\xi+i\zeta^2)^{\alpha}+\eta^*(-i\xi-i\zeta)^{\alpha}]\\
&=|\eta||\xi+\zeta^2|^{\alpha}\cos\left[\frac{\pi\alpha}{2}\sgn{(\xi+\zeta^2)}+\arg{\eta}\right].
\end{split}
\end{equation}
\end{itemize}
In the first two examples, if $-\pi/2\leq(\pi\alpha/2)\sgn(\xi)+\arg{\eta}\leq
\pi/2$, it is easy to show that the sign of $\sigma_{Q}$ remains fixed. 
In the last example, the same is true if 
$-\pi/2\leq(\pi\alpha/2)\sgn(\xi+\zeta^2)+\arg{\eta}\leq\pi/2$. These examples are treated 
exactly in the main body of the paper without resorting to the properties of the
pseudo-differential operators (i.e. the ambiguity resulting from the lack of
knowledge of the exact symbol is circumvented). 

%\bibliography{mybib}
\providecommand{\noopsort}[1]{}\providecommand{\singleletter}[1]{#1}%
\end{document}